\documentclass{egpubl}

\newcommand{\name}{\textsc{PointCleanNet}}
\newcommand{\velodyne}{\textit{Velodyne}}
\newcommand{\kinectone}{\textit{Kinect v1}}
\newcommand{\kinecttwo}{\textit{Kinect v2}}

 \JournalSubmission    
\usepackage[T1]{fontenc}
\usepackage{dfadobe}

\usepackage{cite}
\usepackage[normalem]{ulem}
\usepackage{soul}
\usepackage{color}

\usepackage{xcolor}

\usepackage{amsmath}
\usepackage{amssymb}
\usepackage{bbm}
\usepackage{enumerate}
\usepackage{flushend}
\usepackage{stfloats}

\usepackage{cite}  
\BibtexOrBiblatex
\electronicVersion
\PrintedOrElectronic
\ifpdf \usepackage[pdftex]{graphicx} \pdfcompresslevel=9
\else \usepackage[dvips]{graphicx} \fi

\usepackage{egweblnk}

\newcommand{\mypara}[1]{\paragraph*{#1.}}
\newcommand{\revised}[1]{{\color{black} {#1}}}
\newcommand{\minorrevision}[1]{{\color{black} {#1}}}

\begin{document}
\title [\name: Learning to Denoise and Remove Outliers from Dense Point Clouds]
{\name: Learning to Denoise and Remove Outliers\\from Dense
  Point Clouds}

\author[MJ. Rakotosaona et al.]{
Marie-Julie Rakotosaona$^1$
\hspace{20pt}
Vittorio La Barbera$^2$
\hspace{20pt}
Paul Guerrero$^2$
\hspace{20pt}
Niloy J. Mitra$^2$
\hspace{20pt}
Maks Ovsjanikov$^1$
\\
$^1$LIX, \'{E}cole Polytechnique, CNRS
\hspace{20pt}
$^2$University College London
}

 \maketitle

\begin{figure*}[t!]
    \centering
    \includegraphics[width=\textwidth]{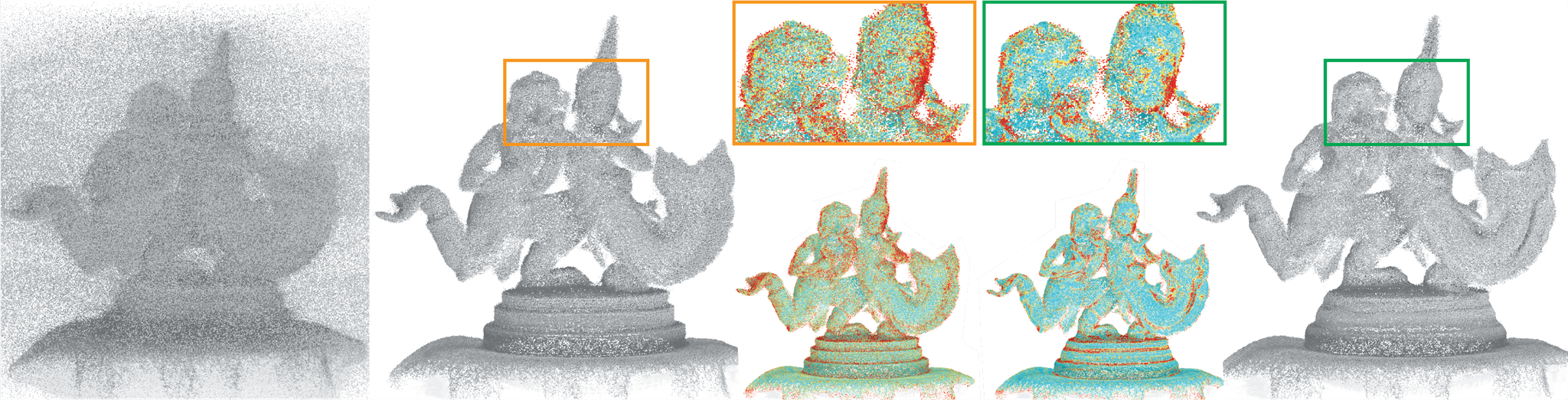}
   \caption{We present \name, a two-stage network that takes a raw point cloud~(left) and first removes
     outliers~(middle) and then denoises the remaining pointset~(right). Our method, unlike many traditional approaches,
     is parameter-free and automatically discovers and preserves high-curvature features without requiring additional
     information about the underlying surface type or device characteristics. Here, point clouds are colored based on
     error  compared to the ground truth point cloud (blue denoting low error, red denoting high error).
 \label{fig:teaser}
\vspace{-2mm}}
\end{figure*}


\begin{abstract}

Point clouds obtained with 3D scanners or by image-based reconstruction techniques are often
corrupted with significant amount of noise and outliers. Traditional methods for point cloud denoising largely rely on local surface fitting (e.g., jets or MLS surfaces), local or non-local averaging, or on statistical assumptions about the underlying noise model. In contrast, we develop a simple data-driven method for removing outliers and reducing noise in unordered point clouds. We base our approach on a deep learning architecture adapted from PCPNet, which was recently proposed for estimating local 3D shape properties in point clouds. Our method first classifies and discards outlier samples, and then  estimates correction vectors that project noisy points onto the original clean surfaces. The approach is efficient and robust to varying amounts of noise and outliers, while being able to handle large densely-sampled point clouds. In our extensive evaluation, both on synthesic and real data, we show an increased robustness to strong noise levels compared to various state-of-the-art methods, enabling accurate surface reconstruction from extremely noisy real data obtained by range scans. 
Finally, the simplicity and universality of our approach makes it very easy to integrate in any existing geometry processing pipeline. \revised{Both the code and pre-trained networks can be found on the project page\footnotemark.}



\end{abstract}


\footnotetext{\revised{https://github.com/mrakotosaon/pointcleannet}}

\section{Introduction}

Raw 3D point clouds obtained directly from acquisition devices such as laser scanners or as output of a reconstruction algorithm (e.g., image-based reconstruction) are regularly contaminated with noise and outliers. The first stage of most geometry processing workflows typically involves \textit{cleaning} such raw point clouds by discarding the outlier samples and denoising the remaining points to reveal the (unknown) scanned surface. The clean output is then used for a range of applications like surface reconstruction, shape matching, model retrieval, etc.

Any good point cloud cleanup algorithm should (i)~\textit{ balance between denoising and feature-preservation}, i.e., remove outliers and noise while retaining data fidelity by preserving sharp edges and local details of the underlying scanned surface; (ii)~\textit{be self-tuning}, i.e., not require as input precise estimates of the noise model or statistics of the unknown scanned surface (e.g., local surface type or curvature characteristics); (iii)~\textit{be invariant to permutation and rigid transform applied to the pointset}, i.e., the denoised output should not depend on angle of scanning or choice of coordinate system; and (iv)~\textit{avoid unnecessarily degrading the input}, i.e., leave the points on the scanned surface if the input happens to be noise-free.  Note that the last criterion implies that the algorithm should not oversmooth the output if the algorithm is iterated multiple times.

Decades of research have produced many variants of denoising approaches targeted for different surface types and noise models (see survey~\cite{han2017review}). Such approaches can be broadly categorized as: classifying points as outliers using statistical methods, e.g., \cite{aggarwal2015outlier}; 
projecting points to estimated local surfaces (e.g., MLS surface, jet-fitting, etc.) \cite{fleishman2005robust,cazals2005estimating,cazals2007jet_fitting_3}; consolidating similar patches to cancel out iid noise perturbations (e.g., non-local means, dictionary-based sparse coding), e.g.,~\cite{digne2012similarity}; or, local smoothing using auxiliary input information (e.g., bilateral smoothing) \cite{huang2013edge}, among many others. Unfortunately, there is no single winner among these methods. The choice of algorithm and its parameters often depends on the scanned surface and the noise characteristics of the acquisition setup. 
Given that the complexity of the underlying geometry and the noise characteristics are, at best, partially known at acquisition time, choosing an optimal algorithm with associated parameters is typically an iterative trial-and-error process.

Inspired by the recent successes  of applying deep learning techniques for the analysis and processing of geometric data, including \cite{masci2015geodesic,bronstein2017geometric,wei2016dense}  among many others, and especially the seminal works designed for learning directly on point clouds \cite{QiEtAl:Pointnet:CVPR:2017,WangEtAl:DGCNN:arxiv:2018}, in this paper, we present \name, a simple data-driven denoising approach. Specifically, we design a two stage point cloud cleaning network based on the recently proposed PCPNet architecture~\cite{GuerreroEtAl:PCPNet:EG:2018} to estimate robust local features and use this information to denoise the point cloud. At training time, a variety of surface patches extracted from a set of shapes is synthetically corrupted with outliers and noise of varying magnitudes (including zero noise). This artificially corrupted set is then used to train \name. Our two-stage method first removes outlier samples and then estimates correction vectors for the remaining points. \revised{Figure~\ref{fig:teaser} shows an example on a raw real-world scanned point cloud from the ETHZ dataset \cite{wolff2016point}.}

The process is enabled by a novel loss function that effectively cleans pointsets without requiring explicit information about the underlying surface or noise characteristics. Intuitively, the network learns to identify local noise-free patches based on estimated features extracted from corresponding raw pointsets and proposes per-point correction vectors. In other words, the network implicitly builds a dictionary of local surface patches in the form of local learned features and uses it to classify input points as outliers and project the remaining ones onto an ensemble of dictionary patches.  At test time, our denoising network directly consumes raw input point clouds, classifies and discards outlier measurements, and denoises the remaining points.  The approach is simple to train and use, and does not expect the user to provide parameters to characterize the surface or noise model. Additionally, unlike traditional approaches, our denoising network can easily be adapted to particular shape families and non-standard noise models. \minorrevision{Similarly to other data-driven techniques, our framework is based on paired noisy-clean data, which we generate synthetically during training, and is thus unlikely to succeed for significantly different test noise.}

We qualitatively and quantitatively evaluate \name\ on a range of synthetic datasets (with access to groundtruth surfaces) and real world datasets.  In our extensive tests, our approach performed better than a variety of state-of-the-art denoising approaches (even with manually-tuned parameters) across both shape and medium to high noise variations. Additionally, the simplicity and universality of our approach makes it very easy to integrate in any existing geometry processing workflow.

\section{Related Work}

Point cloud denoising and outlier removal have a long and rich history in diverse areas of computer
science and a full overview is beyond the scope of the current article. Below, we briefly
review the main general trends for addressing these problems, while concentrating on solutions most
closely related to ours, and refer the interested reader to a recent survey \cite{han2017review}.

\mypara{Outlier removal} The earliest classical approaches for outlier detection, classification
and removal have been proposed primarily in the statistics and data mining communities, in the
general setting of point clouds in arbitrary dimensions, with several monographs dedicated
specifically to this topic
\cite{pincus1995barnett,ben2005outlier,rousseeuw2011robust,aggarwal2015outlier}. These methods are
typically based on robust local statistics and most often come with rigorous theoretical
guarantees. At the same time, their generality often comes at a cost, as purely statistical methods
are often not adapted to the specific features found in geometric 3D shapes, and in most cases
require non-trivial parameter tuning.

More recently, several approaches have been proposed for outlier detection, with emphasis on utility
for 3D point clouds, arising e.g., from acquisition data, including
\cite{chazal2011geometric,guibas2013witnessed,wolff2016point}. The two former methods are
implemented in widely used libraries such as CGAL and have also been used in the context of surface
reconstruction from noisy point clouds \cite{giraudot2013noise}. These approaches are very robust,
but are also based on setting critical parameters or rely on using additional information such as
color \cite{wolff2016point}. This makes it difficult to apply them, for example, across general
noise models, without additional user input and tuning of parameters.

\mypara{Local surface fitting, bilateral filtering} Denoising and outlier removal also arise
prominently, and have therefore been considered in the context of surface fitting to noisy point
clouds, including the widely-used Moving Least Squares (MLS) approach and its robust variants
\cite{alexa2003computing,mederos2003point,fleishman2005robust,oztireli2009feature,gross2011point}. Similarly,
other local fitting approaches have also been used for point cloud denoising, using robust
jet-fitting with reprojection \cite{cazals2005estimating,cazals2007jet_fitting_3} or various forms
of bilateral filtering on point clouds \cite{huang2013edge,digne2017bilateral}, which take into
account both point coordinates and normal directions for better preservation of edge features. A
closely related set of techniques is based on sparse representation of the point normals for better
feature preservation \cite{avron2010,sun2015denoising,mattei2017point}. 
Denoising is then achieved by projecting the points onto the estimated local surfaces. 
These techniques are very
robust for small noise but can lead to significant over smoothing or over-sharpening for high noise
levels \cite{mattei2017point,han2017review}.

\mypara{Non-local means, dictionary-based methods} Another very prominent category of methods,
inspired in part from image-based techniques consist in using non-local filtering based most often
on detecting similar shape parts (patches) and consolidating them into a coherent noise-free point
cloud \cite{deschaud2010point,zheng2010non,digne2012similarity,digne2014self,zeng20183d}. Closely
related are also methods, based on constructing ``dictionaries'' of shapes and their
parts, which can then be used for denoising and point cloud filtering,
e.g., \cite{yoon2016geometry,digne2018sparse} (see also a recent survey of dictionary-based methods
\cite{lescoat2018survey}). Such approaches are particularly well-suited for feature-preserving
filtering and avoid excessive smoothing common to local methods. At the same time, they also require
careful parameter setting and, as we show below, are difficult to apply across a wide variety of
point cloud noise and artefacts.

\mypara{Denoising in images} 
Denoising has also been studied in depth  in other domains such as for images, with a wide variety
of techniques based on both local filtering, total variation smoothing and non-local including
dictionary-based methods
\cite{buades2005non,elad2006image,chambolle2010introduction,mairal2010sparse,elad2010exact}. 

More recently, to address the limitations mentioned above, and inspired by the success of deep
learning for other tasks, several learning-based denoising methods have also been proposed for both
images \cite{zhang2017beyond,zhang2017learning,jin2017deep} \revised{and more recently meshes
  \cite{wang2016mesh,benouini2018efficient},} among others. These methods are especially attractive,
since rather than relying on setting parameters, they allow the method to learn the correct model
from data and adapt for the correct noise setting at test time, without any user
intervention. \revised{In signal processing literature, it is widely believed that image denoising
  has reached close to optimal performance \cite{chatterjee2010denoising,levin2011natural}.}  One of
our main motivations is therefore to show the applicability of this general idea, \revised{and
  especaially the supervised approaches such as \cite{wang2016mesh}, that learn them from a set of
  noisy meshes and their ground-truth counterparts}, to the setting of 3D point clouds.

\mypara{Learning in Point Clouds} Learning-based approaches, and especially those based on deep
learning, have recently attracted a lot of attention in the context of Geometric Data Analysis, with
several methods proposed specifically to handle point cloud data, including PointNet
\cite{QiEtAl:Pointnet:CVPR:2017} and several extensions such as PointNet++
\cite{QiEtAl:Pointnet++:NIPS:2017} and Dynamic Graph CNNs \cite{WangEtAl:DGCNN:arxiv:2018} for shape
segmentation and classification, PCPNet \cite{GuerreroEtAl:PCPNet:EG:2018} for normal and curvature
estimation, P2P-Net \cite{Yin:2018:PBP:3197517.3201288} and PU-Net \cite{yu2018pu} for cross-domain
point cloud transformation and upsampling respectively. Other, convolution-based architectures have
also been used for point-based filtering, including most prominently the recent PointProNet
architecture \cite{RoveriEtAl:PointProNets:CGF:2018}, designed for consolidating input patches,
represented via height maps with respect to a local frame, into a single clean point set, which can
be used for surface reconstruction. Although such an approach has the advantage of leveraging image-based denoising solutions, error creeps in in the local normal estimation stage, especially in the presence of noise and outliers. 

Unlike these techniques, our goal is to train a general-purpose method for removing outliers and
denoising point clouds, corrupted with potentially very high levels of structured noise. For this,
inspired by the success of PCPNet \cite{GuerreroEtAl:PCPNet:EG:2018} for normal and curvature
estimation, we propose a simple framework aimed at learning to both classify outliers and to
displace noisy point clouds by applying an adapted architecture to point cloud patches. We show
through extensive experimental evaluation that our approach can handle a wide range of artefacts,
while being applicable to dense point clouds, without any user intervention.

\section{Overview}
As a first step in digitizing a 3D object, we usually obtain a set of approximate point samples of the scanned surfaces. This \textit{point cloud} is typically an intermediate result used for further processing, for example to reconstruct a mesh or to analyze properties of the scanned object. The quality of these downstream applications depends heavily on the quality of the point cloud. In real-world scans, however, the point cloud is usually degraded by an unknown amount of outliers and noise. We assume the following point cloud formation model:
\begin{equation}
\label{eq:pc_formation}
    \mathbb{P}' = \{p'_i\} =  \{p_i + n_i\}_{p_i \in \mathbb{P}}\ \cup\  
    \{o_j\}_{o_j \in \mathbb{O}},
\end{equation}
where $\mathbb{P}'$ is the observed noisy point cloud, $\mathbb{P}$ are perfect surface samples (i.e., $p_i \in \mathcal{S}$ lying on the scanned surface $\mathcal{S}$), $n_i$ is additive noise, and $\mathbb{O}$ is the set of outlier points. We do not make any assumptions about the noise model $n$ or the outlier model $\mathbb{O}$.
The goal of our work is to take the low-quality point cloud $\mathbb{P}'$ as input, and output a higher quality point cloud closer to $\mathbb{P}$, that is better suited for further processing. We refer to this process as \textit{cleaning}. We split the cleaning into two steps: first we remove outliers, followed by an esimation of per-point displacement vectors that denoise the remaining points:
\begin{equation}
    \tilde{\mathbb{P}} = \{p'_i + d_i\}_{p'_i \in \mathbb{P}'\setminus\tilde{\mathbb{O}}},
\end{equation}
where $\tilde{\mathbb{P}}$ is the output point cloud, $d$ are the displacement vectors and $\tilde{\mathbb{O}}$ the outliers estimated by our method. We first discuss our design choices regarding the desirable properties of the resulting point cloud and then how we  achieve them.

\paragraph*{Approach.}
Traditional statistical scan cleaning approaches typically make assumptions about the scanned surfaces or the noise model, which need to be manually tuned by the user to fit a given setting. This precludes the use of these methods by non-expert users or in casual settings. One desirable property of any cleaning approach is therefore robustness to a wide range of conditions without the need for manual parameter tuning. Recently, deep learning approaches applied to point clouds~\cite{QiEtAl:Pointnet:CVPR:2017, QiEtAl:Pointnet++:NIPS:2017, WangEtAl:DGCNN:arxiv:2018, GuerreroEtAl:PCPNet:EG:2018} have shown a remarkable increase in robustness compared to earlier hand-crafted approaches.

Most of these methods perform a \textit{global} analysis of the point cloud and produce output that depends on the whole point cloud. \revised{This is necessary for global properties such as the semantic class, but is less suited for tasks that only depend on local neighborhoods; processing the entire point cloud simultaneously is a more challenging problem, since the network needs to handle a much larger variety of shapes compared to working with small local patches,
requiring more training shapes and more network capacity. Additionally, processing \emph{dense} point clouds becomes more difficult, due to high memory complexity. }In settings such as ours, local methods such as PCPNet~\cite{GuerreroEtAl:PCPNet:EG:2018} perform better. Both steps of our approach are based on the network architecture described in this method, due to its relative simplicity and competitive performance. We adapt this architecture to our setting (Section~\ref{sec:method}) and train it to perform outlier classification and denoising.

While our cleaning task is mainly a local problem, the estimated displacement vectors $d$ need to be consistent across neighborhoods in order to achieve a smooth surface. With a local approach such as PCPNet, each local estimate is computed separately based on a different local patch. The difference in local neighborhoods causes inconsistencies between neighboring estimates that can be seen as \textit{residual noise} in the result (see Figure~\ref{fig:inconsistent}). We therefore need a method to coordinate neighboring results. We observed that the amount of difference in local neighborhoods between neighboring estimates correlates with the noise model. Thus, the resulting residual noise has a similar noise model as the original noise, but with a smaller magnitude. This means we can iterate our network on the residual noise to keep improving our estimates. See Figure~\ref{fig:architecture} for an overview of the full denoising approach. We will provide extensive experiments with different numbers of denoising iterations in Section~\ref{sec:results}.

\paragraph*{Desirable properties of a point cloud.}
The two stages (i.e., outlier classification and denoising) of our method use different loss functions. The properties of our denoised point cloud are largely determined by these loss functions. Thus, we need to design them such that their optimium is a point cloud that has all desirable properties. We identify two key desirable properties: First, \textit{all points should be as close as possible to the original scanned surface}. Second, \textit{the points should be distributed as regularly as possible on the surface}.
Note that we do not want the denoised points to exactly undo the additive noise and approximate the original perfect surface samples, since the component of the additive noise that is tangent to the surface cannot be recovered from the noisy point cloud.

Section~\ref{sec:training} describes our loss functions, and in Section~\ref{sec:results}, we compare several alternative loss functions.

\section{Cleaning Model}
\label{sec:method}

\begin{figure*}[t!]
    \centering
     \includegraphics[width=\textwidth]{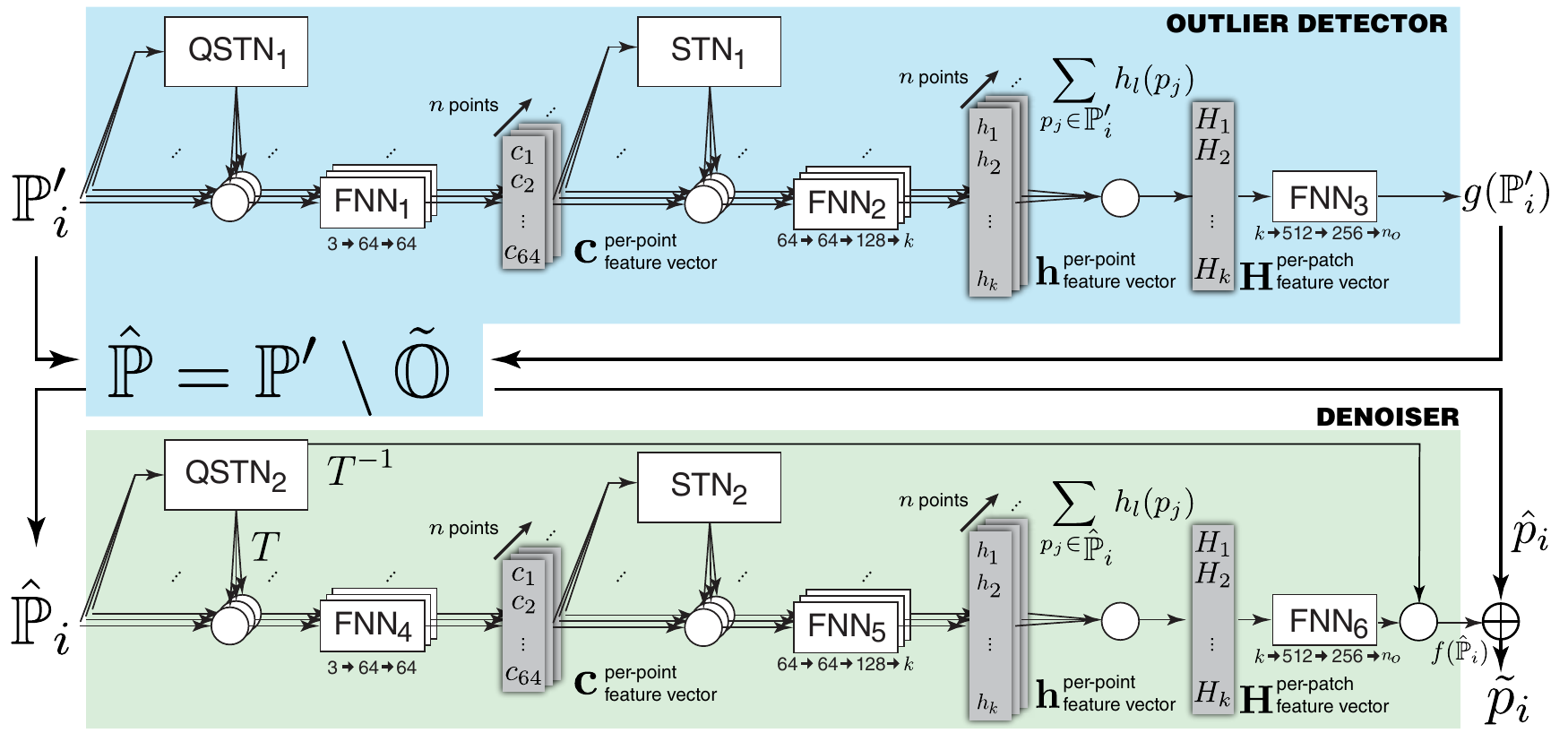}
    \caption{\revised{Our two-stage point cloud cleaning architecture. (Top) Given a noisy point cloud $\mathbb{P}'$, we first apply a local outlier detection network that uses an architecture based on PointNet \cite{QiEtAl:Pointnet:CVPR:2017} and PCPNet \cite{GuerreroEtAl:PCPNet:EG:2018} to detect and remove outliers to obtain $\hat{\mathbb{P}}$ (bottom). We then apply a second network, with a similar architecture, but a different loss, aimed at reducing the noise level in $\hat{\mathbb{P}}$ by estimating correcting displacement vectors, which results in the denoised point cloud $\tilde{\mathbb{P}}$. FNN and (Q)STN stand for fully connected and (Quaternion) Spatial Transformer networks\cite{JaderbergEtAl:STN:NIPS:2015}, similar to their definition and use in PCPNet \cite{GuerreroEtAl:PCPNet:EG:2018}.\vspace{-3mm}}}
    \label{fig:architecture}
\end{figure*}

As mentioned above, our goal is to take a noisy point cloud $\mathbb{P}'$ and produce a cleaned point cloud $\tilde{\mathbb{P}}$ that is closer to the unknown surface that produced the noisy samples. We treat denoising as a local problem: the result for each point $p'_i \in \mathbb{P}'$ only depends on a local neighborhood $\mathbb{P}'_i$ of radius $r$ around the point. Focusing on local neighborhoods allows us to handle dense point clouds without losing local detail. Increasing the locality (or scale) radius $r$ provides more information about the point cloud, at the cost of reducing the capacity available for local details. Unlike traditional analytic denoising approaches, a single neighborhood setting is robust to a wide range of noise settings, as we will demonstrate in Section~\ref{sec:results}. In all of our experiments we set $r$ to $5\%$ of the point cloud's bounding box diagonal.

We assume the point cloud formation model described in Equation~(\ref{eq:pc_formation}), i.e., the noisy point cloud consists of surface samples with added noise and outliers. We then proceed in two stages: first, we train a non-linear function $g$ that removes outliers:
\begin{equation*}
    \tilde{o}_i = g(\mathbb{P}'_i),
\end{equation*}
where $\tilde{o}_i$ is the estimated probability that point $p'_i$ is an outlier. We add a point to the set of estimated outliers $\tilde{\mathbb{O}}$ if $\tilde{o}_i > 0.5$. After removing the outliers, we obtain the point cloud $\hat{\mathbb{P}} = \mathbb{P}'\setminus\tilde{\mathbb{O}}$. We proceed by defining a function $f$ that estimates displacements for these remaining points to move them closer to the unknown surface:
\begin{equation*}
    d_i = f(\hat{\mathbb{P}}_i).
\end{equation*}
The final denoised points are obtained by adding the estimated displacements to the remaining noisy points: $\tilde{p}_i = \hat{p}_i + d_i$.
Both $f$ and $g$ are modeled as deep neural networks with a PCPNet-based architecture. We next provide a short overview of PCPNet before describing our modifications.

A major challenge when applying deep learning methods directly to point clouds is achieving invariance to the permutation of the points: all permutations should produce the same result. Training a network to learn this invariance is difficult due to the exponential number of such permutations. As a solution to this problem, PointNet~\cite{QiEtAl:Pointnet:CVPR:2017} proposes a network architecture that is order-invariant by design. However, PointNet is a global method, processing the whole point cloud in one forward iteration of the network. This results in a degraded performance for shape details. PCPNet~\cite{GuerreroEtAl:PCPNet:EG:2018} was proposed as a local variant of PointNet that is applied to local patches, gives better results for shape details, and is applicable to dense point clouds, possibly containing millions of points. We base our denoising architecture on PCPNet.

\paragraph*{Creating a local patch.}
Given a point cloud $\mathbb{P} = \{p_1, \dots, p_n\}$, the local patch $\mathbb{P}_i$ contains all the points within the constant radius $r$ inside a ball centered around $p_i$. Using this patch as input, we want to compute the outlier probability $\tilde{o}_i$ and a displacement vector $d_i$, for the remaining non-outlier points. We first normalize this patch by centering it and scaling it to unit size. The PCPNet architecture requires patches to have a fixed number of points; like in the original paper, we pad patches with too few points with zeros and take a random subset from patches with too many points. Intuitively, this step, makes the network more robust to additional points.

\paragraph*{Network architecture.}
An overview of our network architecture is shown in Figure~\ref{fig:architecture}. Given the normalized local patch $\mathbb{P}_i$, the network first applies a spatial transformer network~\cite{JaderbergEtAl:STN:NIPS:2015} that is constrained to rotations, called a quaternion spatial transformer network (QSTN). This is a small sub-network that learns to rotate the patch to a canonical orientation (note that this estimation implicitly learns to be robust to outliers and noise, similar to robust statistical estimation). \revised{At the end of the pipeline, the final estimated displacement vectors are rotated back from the canonical orientation to world space.} The remainder of the network is divided into three main parts:
\begin{itemize}
    \item a \textit{feature extractor} $h(p)$ that is applied to each point in the patch separately,
    \item a \textit{symmetric operation} $H(\mathbb{P}_i) = \sum_{p_j \in \mathbb{P}_i} h(p_j)$ that combines the computed features for each point into an order-invariant feature vector for the patch, and
    \item a \textit{regressor} that estimates the desired properties $d_i$ and $\tilde{o}_i$ from the feature vector of the patch.
\end{itemize}
Following the original design of PointNet \cite{QiEtAl:Pointnet:CVPR:2017}, the feature extractor is implemented with a multi-layer perceptron that is applied to each point separately, but shares weights between points. Computing the features separately for each point ensures that they are invariant to the ordering of the points. The feature extractor also applies an additional spatial transformer network to intermediate point features. 

In our implementation, we add skip connections to the multi-layer perceptrons, similar to ResNet blocks.
Empirically, we found this to help with gradient propagation and improve training performance.
The regressor is also implemented with a multi-layer perceptron. Similar to the feature extractor, we add skip connections to help gradient propagation and improve training performance. We use the same network width as in the original PCPNet (please refer to the original paper for details). However, the network is two times deeper as we replace the original layers with two layers ResBlocks.
This architecture is used to compute both outlier indicators and displacement vectors. We change the number of channels of the last regressor layer to fit the size of the desired output ($1$ for outlier indicators and $3$ for displacement vectors).

Importantly, for a each point $p_i$ in the point cloud, we compute its local neighborhood $\mathbb{P}_i$ and only estimate the outlier probability and displacement vector for the center point $p_i$, i.e., we do not estimate outlier probabilities or displacement vectors for other points in the patch. Thus, each point in the original point cloud is processed independently by considering its own neighborhood and indirectly gets coupled by the iterative cleaning, as described next.

\begin{figure}[b!]
    \centering
    \includegraphics[width=\columnwidth]{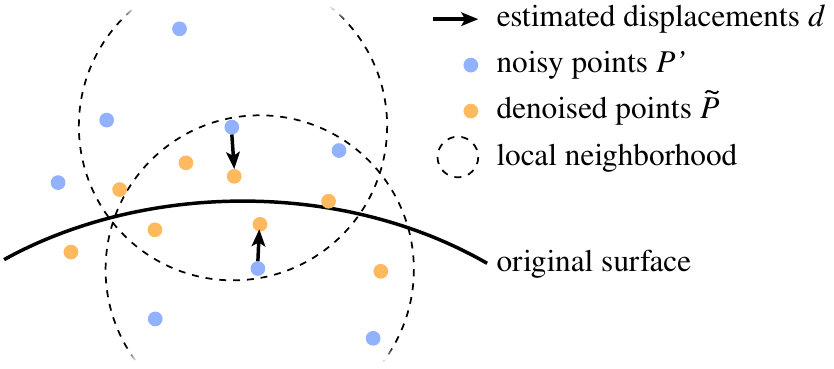}
    \caption{Residual noise. Different local neighborhoods for adjacent denoised points cause slightly different results, which can be seen as residual noise in the denoised points. Iterating the denoising approach improves the results.}
    \label{fig:inconsistent}
\end{figure}

\paragraph*{Iterative cleaning.}
At test time, after applying the displacement vectors computed from a single iteration of the architecture, we are left with residual noise. The residual error vectors from denoised points $\tilde{p}_i$ to the target surface that are introduced by our method do not vary smoothly over the denoised points. \revised{Empirically, we found that this residual noise has a similar type, but a lower magnitude than the original noisy points. Intuitively, this can be explained by looking at the content of input patches for points that are neighbors in the denoised point cloud. As shown in Figure~\ref{fig:inconsistent}, input patches that are far apart have different content, resulting in different network predictions, while patches that are close together have similar content, and similar predictions. The distance of these input patches correlates with the noise model and the noise magnitude, therefore the network predictions, and the denoised points, are likely to have noise of a similar type, but a lower magnitude than the original noisy points.
This allows us to iterate our denoising approach to continue improving the denoised points.}

In practice, we observed shrinking of the point cloud after several iterations. To counteract this shrinking, we apply an inflation step after each iteration, inspired by Taubin smoothing~\cite{Taubin:TS:CV:1995}:
\begin{equation}
    d'_i = d_i - 1/k \sum _{p_j \in N(p_i)} d_j,
\end{equation}
where $d'_i$ are the corrected displacements vectors and $N(p_i)$ are the $k$ nearest neighbours of point $p_i$, we set $k=100$. Note that this step approximately removes the low-frequency component from the estimated displacements.

\section{Training Setup}
\label{sec:training}

To train the denoising model, we use a dataset of paired noisy point clouds and corresponding clean ground truth point clouds. We do not need to know the exact correspondences of points in a pair, but we assume we do know the ground truth outlier label for each noisy point. Using a point cloud as ground truth instead of a surface description makes it easier to obtain training data. For example, a ground truth point cloud can be obtained from a higher-quality scan of the same scene the noisy point cloud was obtained from. Since we work with local patches instead of entire point clouds, we can train with relatively few shapes. To handle different noise magnitudes, and to enable our iterative denoising approach, we train with multiple noise levels. This includes several training examples with zero noise magnitude, which trains the network to preserve the shape of point clouds without noise.

\mypara{Loss function} Choosing a good loss function is critical, as this choice has direct impact on the properties of the cleaned point clouds. For the outlier removal phase, we use the $L_1$ distance between the estimated outlier labels and the  ground truth outlier labels:
\begin{equation}
    L_o(\tilde{p}_i, p_i) = \|\tilde{o}_i - o_i\|_1,
\end{equation}
where $\tilde{o}_i$ is the estimated outlier probability and $o_i$ is the ground truth label. We also experimented with the binary cross-entropy loss, but found the $L_1$ loss to perform better, in practice.

\begin{figure}[t!]
    \centering
    \includegraphics[width=.8\columnwidth]{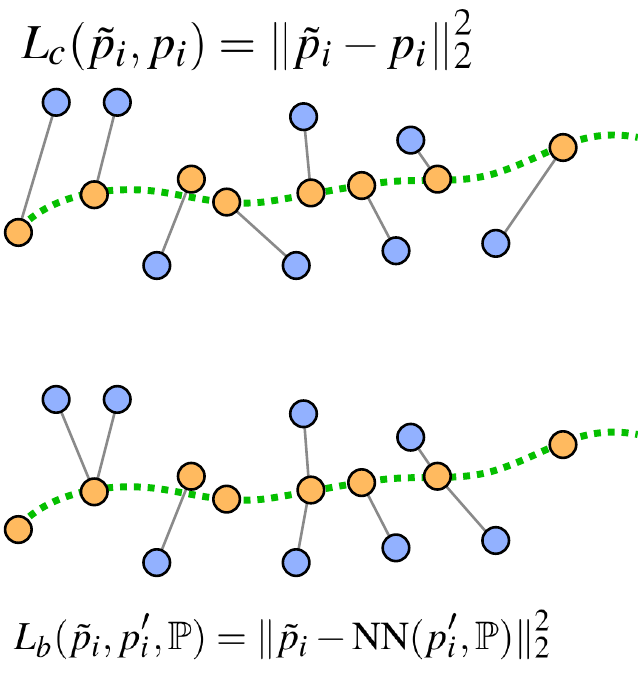}
    \caption{Alternate loss functions that result in comparatively worse performance (see Figure~\protect\ref{fig:losses}). Dotted green line denotes the  underlying scanned surface, orange points denote original points, and blue points denote the noisy points.  The error function $L_c$ (top) tries to learn denoising as denoised points $\tilde{p_i}$ going back to the original ground truth points ${p_i}$; while, the error function $L_b$ tries to learn denoising as denoised points $\tilde{p_i}$ going  to the closest point in the cleaned point set $\mathbb{P}$, 
    i.e., NN$(p'_i, \mathbb{P})$. 
    }
    \label{fig:alternate_loss}
\end{figure}

In the denoising setting, designing the loss function is less straight-forward. Two properties we would like our denoised point clouds to have are proximity of the points to the scanned surface, and a regular distribution of the points over the surface. Assuming the ground truth point cloud has both of these properties, a straight-forward choice for the loss would be the $L2$ distance between the cleaned and the ground truth point cloud:
\begin{equation}
    L_c(\tilde{p}_i, p_i) = \|\tilde{p}_i - p_i\|^2_2,
\end{equation}
where $\tilde{p}_i$ and $p_i$ are corresponding cleaned and ground truth points in a patch. Note that, for simplicity of notation, we have unrolled the displacement vector expressions directly in terms of point coordinates. However, this assumes knowledge of a point-wise correspondence between the point clouds; and even if the correspondence is known, we can in general not recover the component of the additive noise that is tangent to the surface.
The minimizer of this loss is therefore an average between all potential candidates the noisy point may have originated from. This average will in general not lie on the surface, and lead to poor overall performance. Figure~\ref{fig:alternate_loss}, top, illustrates this baseline loss. Fortunately, we do not need to exactly undo the additive noise. There is a large space of possible point clouds that satisfy the desired properties to the same degree as the ground truth point cloud, or even more so.

\begin{figure}[t!]
\centering
  \includegraphics[width=\columnwidth]{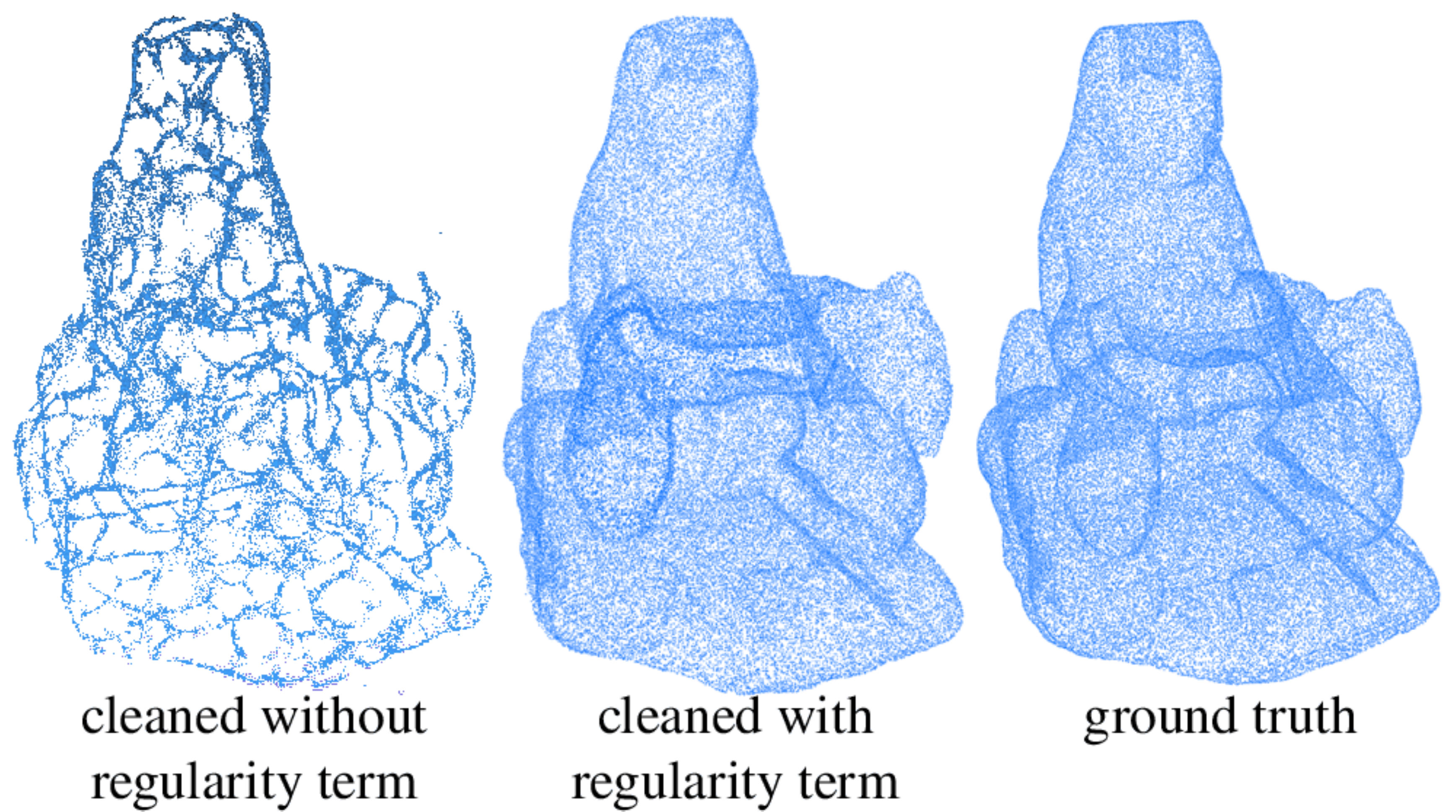}
  \caption{Omitting the regularity term from the loss causes the points to cluster into filament structures on the surface after multiple iterations (left). Compare to results with the regularity term (center) and the ground truth point cloud (right).
  \label{fig:filaments}
\vspace{-2mm}
}
\end{figure}

We propose a main loss function and an alternative with a slightly inferior performance, but simpler and more efficient formulation. The main loss function has one term for each of the two properties we would like to achieve: \textit{Proximity to the surface} can be approximated as the distance of each denoised point to its nearest neighbour in the ground truth point cloud:
\begin{equation}
\label{eq:loss_nearest}
L_s(\tilde{p}_i, \mathbb{P}_{\tilde{p}_i}) = \min_{p_j \in \mathbb{P}_{\tilde{p}_i}} \|\tilde{p}_i - p_j\|^2_2.
\end{equation}
For efficiency, we restrict the nearest neighbor search to the local patch $\mathbb{P}_{{\tilde{p}_i}}$ of ground truth points centered at $\tilde{p}_i$. Originally, we experimented with only this loss function, but noticed a filament structures forming on the surface after several denoising iterations, as shown in Figure~\ref{fig:filaments}. Since the points are only constrained to lie on the surface, there are multiple displacement vectors that bring them equally close to the surface. In multiple iterations, the points drift tangent to the surface, forming clusters. To achieve a more \textit{regular distribution on the surface}, we introduce a regularization term:
\begin{equation}
  \label{eq:loss_farthest}
  L_r(\tilde{p}_i, \mathbb{P}_{\tilde{p}_i}) = \max_{p_j \in \mathbb{P}_{\tilde{p}_i}} \|\tilde{p}_i - p_j\|^2_2.
\end{equation}
\revised{By minimizing this term, we minimize the squared distance to the \emph{farthest} point in the local patch $\mathbb{P}_{\tilde{p}_i}$. Intuitively, this keeps the cleaned point centered in the patch and discourages a drift of the point tangent to the surface. Assuming the noisy point clouds are approximately regularly distributed, this results in a regular distribution of the cleaned points since, in this case Eq. \ref{eq:loss_farthest} promotes the clean point to lie in the barycenter of the points in its patch. With this term, we want to avoid the excessive clustering of points (for example, into filament structures), which is especially important when applying our approach iteratively.} The full loss function is a weighted combination of the two loss terms:
\begin{equation}
  \label{eq:total_loss}
    L_a = \alpha\ L_s + (1-\alpha)\ L_r.
\end{equation}
Since the second term can be seen as a regularization, we set $\alpha$ to $0.99$ in our experiments.

Importantly, the loss defined in Eq. (\ref{eq:total_loss}) depends on the current point cloud, so that the point searches in Equations 
(\ref{eq:loss_nearest}) and (\ref{eq:loss_farthest}) need to be \emph{updated in every training epoch}. Alternatively, these target points can be fixed. Thus, our alternative loss function uses an explicit ground truth for the cleaned point that can be precomputed:
\begin{equation}
  \label{eq:alternative_loss}
    L_b(\tilde{p}_i, p'_i, \mathbb{P}) = \|\tilde{p}_i - \mathrm{NN}(p'_i, \mathbb{P})\|^2_2, 
\end{equation}
where $\mathrm{NN}(p'_i, \mathbb{P})$ is the closest point to the initial noisy point $p'_i$ (before denoising) in the ground truth point set $\mathbb{P}$.
Figure~\ref{fig:alternate_loss}, bottom, illustrates this loss. Since both $p'_i$ and the ground truth point cloud are constant during training, this mapping can be precomputed, making this loss function more efficient and easier to implement. Additionally, the fixed target prevents the points from drifting tangent to the surface. However, this loss constrains the network more than $L_a$ and we observed a slightly lower performance.

For the outlier removal network we use a learning rate of $10^{-4}$ and uniform Kaiming initialization~\cite{He:KI:ICCV:2015} of the network weights. When training the denoising network, we observed that network weights converge to relatively small values. To help convergence, we lower the intial values of the weights to uniform random values in $[-0.001, 0.001]$ and decrease the learning rate to $10^{-8}$. This improves convergence speed for the denoising network and lowers the converged error.

\begin{figure}[t!]
\centering
  \includegraphics[width=\columnwidth]{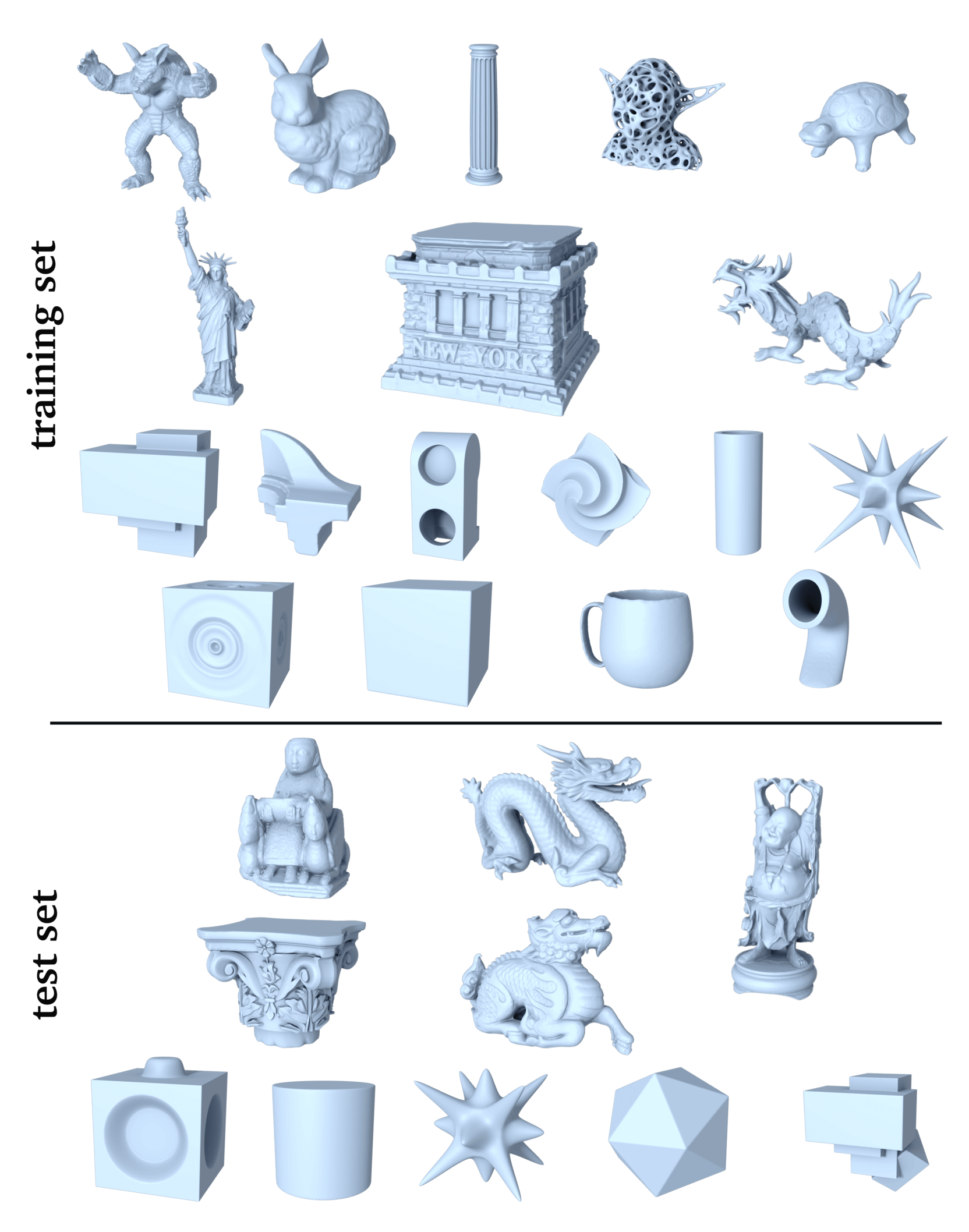}
  \caption{\revised{The shapes used for the \name\ training and test sets.}
  \label{fig:pointcleannet_dataset}
\vspace{-2mm}
}
\end{figure}

\begin{figure*}[t]
    \centering
    \includegraphics[width=\textwidth]{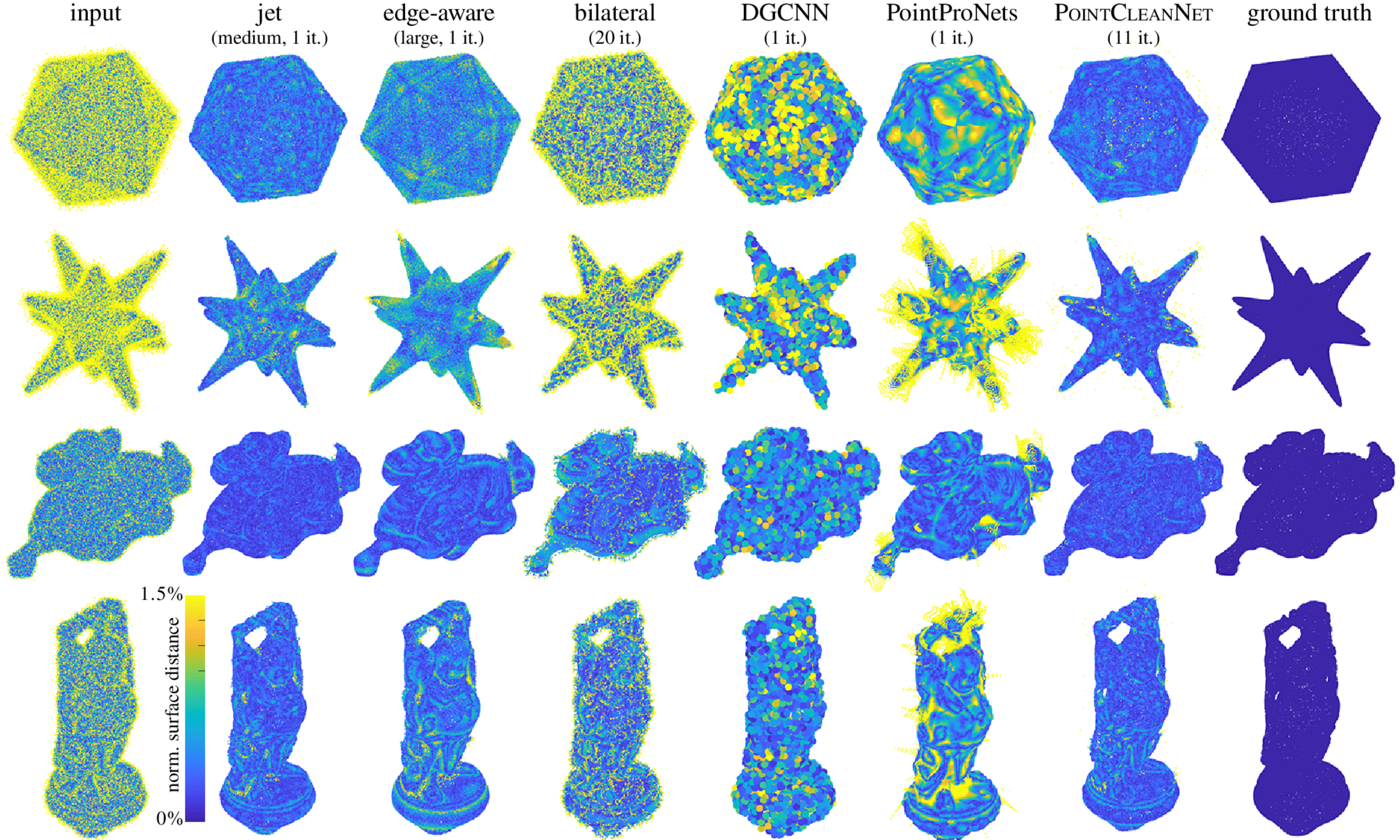}
    \caption{\revised{Qualitative comparison to the state-of-the-art. We compare simple shapes in the top row and increase shape complexity towards the bottom. DGCNN can only handle small point clouds, thus we use a sparser sampling for this method. Colors illustrate the denoising error, we use the distance-to-surface for each denoised point.}
    \label{fig:qual_1}
\vspace{-2mm}}
\end{figure*}

\revised{

  \subsection{Relation to PCPNet}

While being directly based on PCPNet~\cite{GuerreroEtAl:PCPNet:EG:2018}, our approach has several characteristics, specifically adapted to the point cloud denoising and outlier detection problem:

\paragraph*{Loss.} We use an adapted loss, summarized in Eq. (\ref{eq:loss_nearest}) and (\ref{eq:loss_farthest}), which, importantly, not only includes a regularization via the distance to the farthest point, but is also updated at every training iteration, through the change of the corresponding points. We have experimented with several alternatives such as the loss described in Eq. (\ref{eq:alternative_loss}) and found them to perform consistently worse than ours.
\paragraph*{Iterative deep network.} Importantly, we apply our network \emph{iteratively} for improved noise reduction. While perhaps non-standard, this results in very significant improvement in our setting. Moreover, we found that a straightforward implementation might not converge, while with the proper loss and with an inflation term, the network can both stabilize and achieve higher accuracy. 
\paragraph*{Practical applicability.} Finally, we remark that \name, as a specialized adaptation of an existing network, both simplifies its integration in practice and establishes its applicability for point cloud denoising and outlier removal. }

\section{Results}
\label{sec:results}

\revised{We first describe our dataset and evaluation metric in Sections~\ref{sec:dataset} and \ref{sec:eval_metric}. Based on this dataset and metric, we compare the denoising performance (Sec. \ref{sec:results_denoising}) and the outlier detection performance (Sec. \ref{sec:results_outlier_removal}) of our method to several baselines and state-of-the-art methods, including the recent learning-based approaches PointProNets~\cite{RoveriEtAl:PointProNets:CGF:2018} and an adapted version of Dynamic Graph CNNs~\cite{WangEtAl:DGCNN:arxiv:2018} among others. Experiments on additional datasets with different noise distributions, including simulations of non-uniform scanner noise, and noise from real world scans are presented in Section~\ref{sec:noise_distributions}. \minorrevision{We evaluate our method under different forms of point cloud artefacts classified in a recent survey \cite{berger2017survey}. In particular, we treat noisy data and outliers in Sec. \ref{sec:results_denoising} and \ref{sec:results_outlier_removal}, and consider misaligned scans in \ref{sec:misaligned_data}. Although we do not perform an extensive evaluation, we also remark that our method is, to some extent, robust to non-uniform sampling due to the regularization term from Eq.~\eqref{eq:loss_farthest} that promotes regularity in the denoised point cloud.}}

\subsection{Datasets}
\label{sec:dataset}
\revised{Our main dataset contains 28 different shapes, which we split into 18 training shapes and 10 test shapes. See Figure~\ref{fig:pointcleannet_dataset} for a gallery of all shapes. From the original triangle meshes of each shape, we sample 100K points, uniformly at random on the surface, to generate a clean point cloud. 

For the \emph{denoising task}, noisy point clouds are generated by adding Gaussian noise with the standard deviation of 0.25\%, 0.5\%, 1\%, 1.5\% and 2.5\% of the original shape's bounding box diagonal. In total, the denoising training set contains 108 shape variations, arising from 6 levels of noise (including the clean points) for each of the 18 shapes.}

For the \emph{outlier removal task}, we use the same training and test shapes, however we use only clean point clouds and with a larger sample count of 140k points per shape.
\revised{To generate outliers, we added Gaussian noise with standard deviation of 20\% of the shape's bounding box diagonal to a random subset of points. The training set contains point clouds with proportions starting at 10\% until 90\% in intervals of 10\% of the points converted to outliers.}
Only the outliers that are farther from the surface than the standard deviation of the noise distribution are selected. 

In total, the outlier removal training set contains 432 example shapes, arising from 6 outlier densities and 4 levels of noise for each of the 18 training shapes.

The test set contains point clouds with 30\% of outliers points. To test the generality of our outlier removal, we added a second method to generate outliers to our test set only. In this setting, outliers are distributed uniformly inside the shape's bounding box that has been scaled up by 10\%.

\revised{In Section~\ref{sec:noise_distributions} we also evaluate our method on point clouds generated with alternative methods, including simulated non-uniform noise and noise from real real acquisition devices.

\name\ training datasets for denoising and outlier removal are available on our
\href{http://github.com/mrakotosaon/pointcleannet}{project page}.}

\begin{figure*}[t!]
    \centering
    \includegraphics[width=\textwidth]{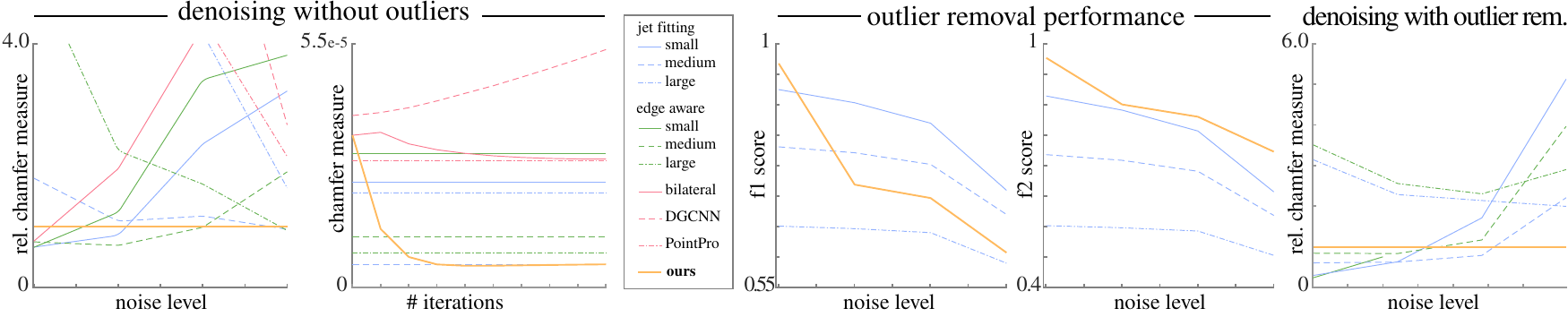}
    \caption{\revised{Quantitative comparison. We compare the performance of our model to jet
        smoothing~\cite{cazals2005estimating}, edge-aware denoising~\cite{huang2013edge}, the bilateral point cloud
        filter by Digne et al.~\cite{digne2017bilateral}, Dynamic Graph CNNs~\cite{WangEtAl:DGCNN:arxiv:2018}, and
        PointProNets~\cite{RoveriEtAl:PointProNets:CGF:2018}. The two plots on the left are evaluated on our test set
        without outliers, the two following plots compare the outlier removal performance using the f1 and the f2 scores, and the right-most plot shows the denoising performance after outlier removal.}
    \label{fig:quant_1}
\vspace{-2mm}}
\end{figure*}

\subsection{Evaluation Metric}
\label{sec:eval_metric}
The evaluation metric should be sensitive to the desired properties of the point cloud described earlier: point clouds should be close to the surface and have an approximately regular distribution. If we assume the ground truth point clouds have a regular distribution, \revised{the following \textit{Chamfer measure}~\cite{fan2017point,achlioptas2018learning}, a variant of the Chamfer distance~\cite{Barrow:1977:PCC}, measures both of these properties}:
\begin{equation*}
    c(\tilde{\mathbb{P}}, \mathbb{P}) = \frac{1}{N} \sum_{p_i \in \tilde{\mathbb{P}}} \min_{p_j \in \mathbb{P}} \|p_i-p_j\|^2_2 + \frac{1}{M} \sum_{p_j \in \mathbb{P}} \min_{p_i \in \tilde{\mathbb{P}}} \|p_j-p_i\|^2_2.
\end{equation*}
\revised{Here $N$ and $M$ are the cardinalities of the cleaned $\tilde{\mathbb{P}}$ and ground truth $\mathbb{P}$ point clouds, respectively. Note that the first term measures an approximate distance from each cleaned point to the target surface, while the second term intuitively rewards an even coverage of the target surface and penalizes gaps. All our point clouds are scale-normalized to have a unit bounding box diagonal, making the point distances comparable for different shapes.

For a dataset with simulated scanner noise that we will describe in Section~\ref{sec:noise_distributions}, the clean point set has a non-uniform point distribution. For this dataset we use only the root mean square distance-to-surface (\emph{RMSD}) of each point as evaluation metric:
\begin{equation*}
    d(\tilde{\mathbb{P}}, \mathbb{P}) = \sqrt{\frac{1}{N} \sum_{p_i \in \tilde{\mathbb{P}}} \min_{p_j \in \mathbb{P}} \|p_i-p_j\|^2_2)}.
\end{equation*}
}

\subsection{Evaluating Denoising}
\label{sec:results_denoising}

\begin{figure*}[t]
    \centering
    \includegraphics[width=\textwidth]{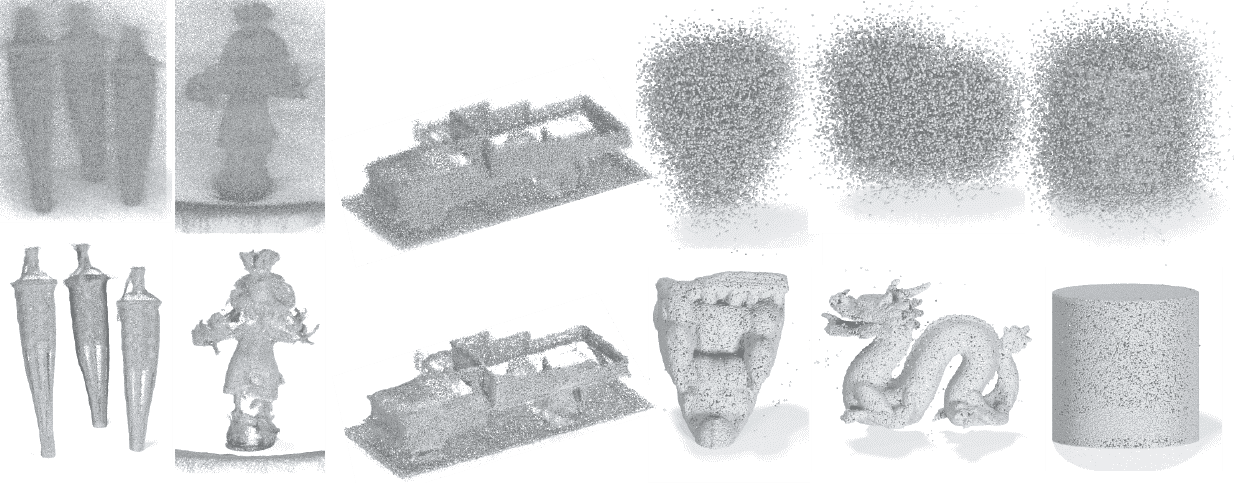}
    \caption{Example input 
    outlier and noise corrupted pointclouds~(top) and their corresponding cleaned output~(bottom) produced by \name. From left to right: torch, scarecrow, tanks-and-temple, galera, dragon, cylinder. \revised{The left two examples are corrupted with real-world scanning noise and outliers, the other examples with synthetic noise and outliers.}}
    \label{fig:results_plate}
\end{figure*}

We first evaluate the denoising task alone, without outlier removal. We compare the results of our method on different noise levels to several state-of-the-art techniques for point cloud denoising.

We first consider a qualitative evaluation of our results in Figure~\ref{fig:qual_1}, showing the denoised point clouds for four different input noisy point clouds (Icosahedron, Star smooth, Netsuke, Happy) with two different noise intensities, 1\% and 2.5\% of the original shape bounding box diagonal. The distances from each of the denoised points to the ground truth surface are color-coded.
In the same figure, we can also compare the performance of our method to other successful algorithms.

\revised{We compare against five other methods, as described next. It is important to note that in most of these methods, it is necessary to tune some parameters, such as the neighborhood size, to adjust to the different noise levels, while our method works across all noise levels with the same hyper-parameters.
When applicable, we manually adjusted parameters for best performance. Also, in some algorithms, we allowed multiple parameter settings (small, medium, large) to handle different levels of noise.

\begin{enumerate}[(i)]
    \item \textit{Polynomial fitting with osculating jets} \cite{cazals2005estimating,cazals2007jet_fitting_3}: Osculating jet-fitting performs well if the right neighborhood size is chosen for the given noise level. Otherwise, neighborhood sizes that are too small overfit to strong noise (Figure~\ref{fig:qual_1}, first row), and neighborhood sizes that are too large do not preserve detailed features (Figure~\ref{fig:qual_1}, second and fourth row). 
    \item \textit{Edge-aware point set resampling} \cite{huang2013edge}: Edge-aware point set resampling has larger errors near detailed features (Figure~\ref{fig:qual_1}, third and fourth row), while obtaining good results near sharp edges, like the edges of the icosahedron.
    \item \textit{Bilateral filtering for point clouds}~\cite{digne2017bilateral}: bilateral filtering performs poorly in strong noise settings (Figure~\ref{fig:qual_1}, first two rows). 
    \item \textit{Dynamic Graph CNN (DGCNN)}~\cite{WangEtAl:DGCNN:arxiv:2018}: Note that Dynamic Graph CNNs were not designed for local operations, such as denoising. We modify the segmentation variant of this method to output a displacement vector per point instead of class probabilities. For the loss, the displacements are added to the original points and the result is compared to the target point cloud using the same Chamfer measure used as the error metric in our evaluation. 

Since the whole point cloud is processed in a single go, we need to heavily sub-sample our dense point clouds before using them as input for DGCNN. We also restrict DGCNN to a single iteration as we found the result set to diverge over iterations. Similar to bilateral filtering, DGCNN also performs poorly in strong noise settings (Figure~\ref{fig:qual_1}, first two rows). 
\item \textit{PointProNets} \cite{RoveriEtAl:PointProNets:CGF:2018}: PointProNets requires oriented normals during training. Where available, these are obtained from the ground truth source meshes, or estimated with PCPNet~\cite{GuerreroEtAl:PCPNet:EG:2018} otherwise. Differently from the original method, we also use \emph{ground truth normals} to orient the predicted height maps, instead of trying to estimate an orientation, as we found the in-network estimation used in the original method to be unstable for our datasets. Note that this provides an upper bound for the performance of PointProNets. The denoised patches do not accurately reconstruct detailed surfaces, presumably due to the smoothing effect of the image convolutions, and suffer from artefacts caused by the smoothing of the height map at the boundaries of a patch.
\end{enumerate}
}

In contrast, our method works on local patches directly in the point domain, can apply several iterations of denoising to improve results, and is robust to a large range of noise levels with the same choice of hyper-parameters. This results in lower residual error, especially in detailed surface regions.

We also present quantitative comparisons that summarize the performance of each method on the entire dataset. The previously described Chamfer measure is used as evaluation metric that captures both the distance from denoised points to the ground truth surface and the regularity of the points.

\revised{Results are shown in Figure ~\ref{fig:quant_1} left (without outlier removal).
We can observe that \name\ performs noticeably better under mid to high noise level and using multiple iterations compared to all the other methods. The performance of our method is also more stable to changes in noise levels, while most other methods perform well only for a specific level of noise.}

\begin{figure}[t]
    \centering
    \includegraphics[width=\columnwidth]{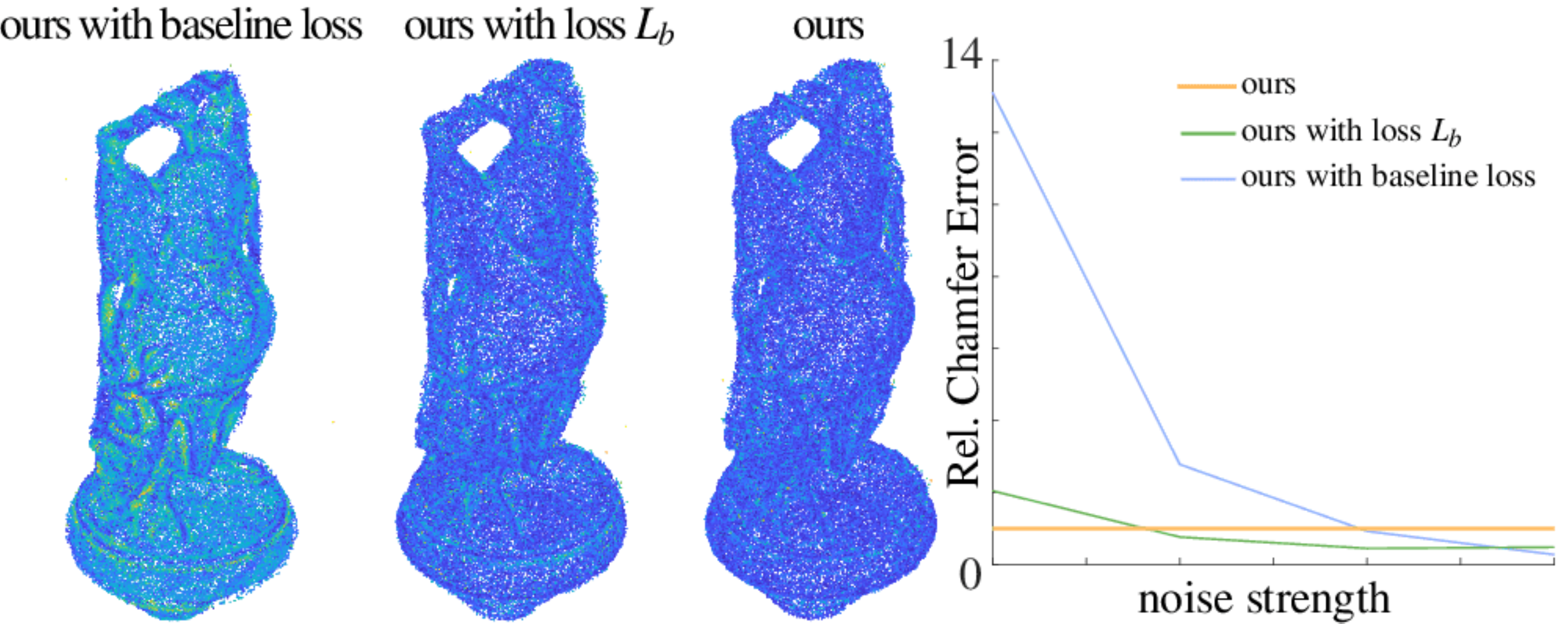}
    \caption{Comparison of our two-term loss $L_a$ with the alternative loss $L_b$ and the baseline loss $L_c$. Our loss results gives large benefits over the baseline loss, and performs somewhat better than the alternative loss as well, due to being less constrained. }
    \label{fig:losses}
\end{figure}

\paragraph*{Comparison to the alternative loss.} As shown in Figure~\ref{fig:losses}, our alternative loss performs slightly worse than our main loss. However, it is more efficient and easier to implement, so the choice of loss function depends on the setting.

\subsection{Evaluating Outlier Removal}
\label{sec:results_outlier_removal}

\begin{figure*}[t]
    \centering
    \includegraphics[width=\textwidth]{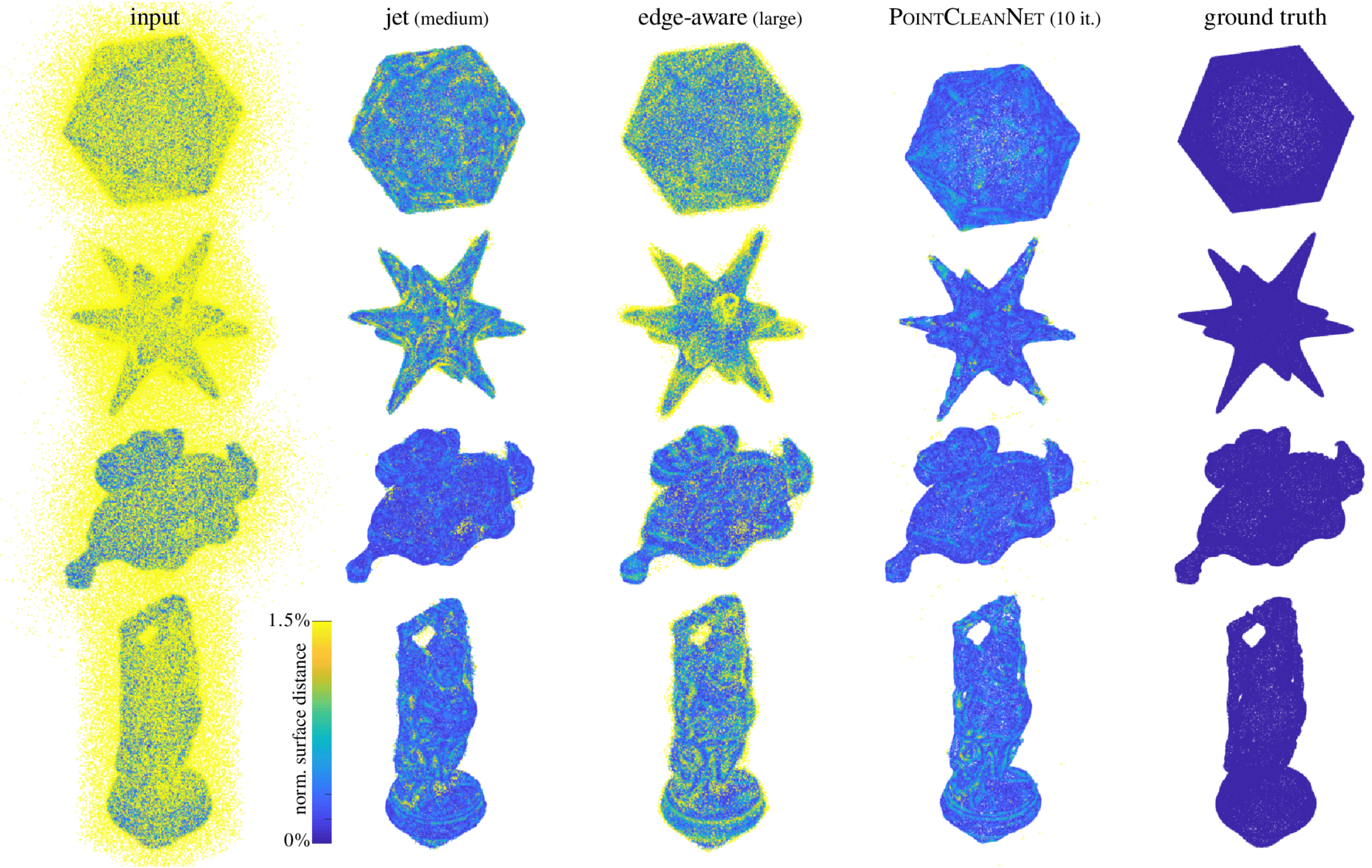}
    \caption{Qualitative comparison on our outlier test set. Here the task is to remove points that are not part of the original surface. Note that that analytic methods with a large setting for the radius (third column) fail to remove outliers hidden inside small details, such as the arms of the statue, while a smaller setting (second column) results in a lot of residual noise. Since \name\ can learn to adapt to the feature to produce a result with less noise.}
    \label{fig:qual_outliers}
\end{figure*}

\revised{Figure \ref{fig:quant_1} shows the performance of our outlier removal method (right). For the purpose of
  cleaning dense data, a model should prioritize classifying outlier points correctly (true positives) over limiting
  the number of false positives. Therefore we consider that recall has more importance than precision for this task. The
  $F_\beta$ score conveys the balance between recall and precision. Plots 3 and 4 compare our method to
  jet-fitting~\cite{cazals2005estimating} using $F_1$ and $F_2$ scores. We observe that when recall and precision are
  weighted equally, our method has the best performance when removing outliers on clean point clouds while remaining
  effective on the other noise levels. \name\ performs the best for all noise levels when using $F_2$ score
  which gives larger weight to recall.

The last plot in Figure \ref{fig:quant_1} compares our approach to jet-fitting and edge-aware
filtering~\cite{huang2013edge} with both outlier removal and denoising on the test set. In this experiment, we first
removed outliers using an outlier classification technique and then denoised the point clouds from our test set. We show the results for different noise levels from zero to 2.5\% of the shape bounding box diagonal. Finally, we make two observations: first, \name\ outperforms edge-aware and jet-fitting techniques with outlier removal and denoising on medium to large noise levels; and second, on smaller noise levels, our model still outperforms a few of the different tuning variations of the related techniques. Recall that our model does not require parameter tuning by the user.}

Figure \ref{fig:qual_outliers} also shows qualitative results for outlier removal on our test set compared to the related techniques mentioned before. We observe that edge-aware filtering  performs worse around highly detailed regions and edges, while jet-fitting  does not manage to clean the remaining outliers at scattered points. The result highlights the consistent performance of \name\ across different shapes and varying level of details, contrary to the other methods which produce less consistent distances to the underlying ground truth shapes.

\subsection{Performance under Different Noise Types}
\label{sec:noise_distributions}

\begin{figure*}[t]
    \centering
    \includegraphics[width=\textwidth]{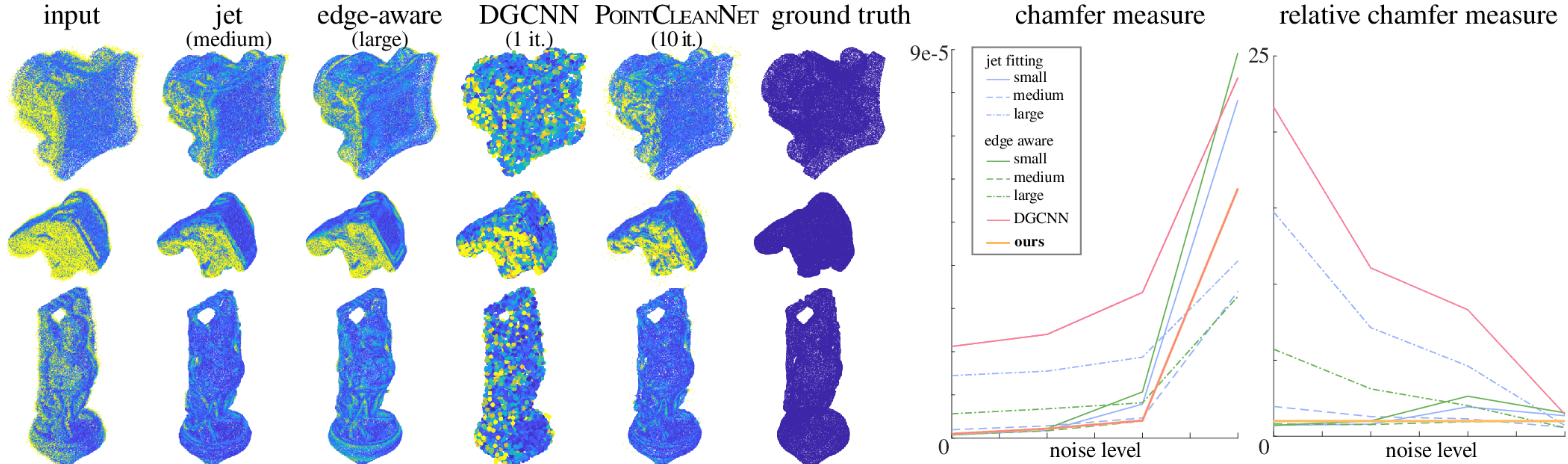}
    \caption{Qualitative comparison on the directional noise test set. We show that \name\ can adapt to different kinds of noise models. Here we added anisotropic noise to the shapes. \revised{Qualitative results are o the left, and quantitative plots of the absolute and relative Chamfer measure over different noise levels on the right. Even though our method was not trained on isotropic noise only, it still performs on par with the state of the art on low and medium noise levels.}}
    \label{fig:dir_noise}
\end{figure*}

\subsubsection{Directional noise}
We evaluated \name\ on a synthetic dataset simulating 3D data acquisition via depth cameras. To do so we created a dataset with structured noise levels to simulate depth uncertainty of depth reconstructions.
Specifically, we added noise using an anisotropic Gaussian distribution with constant covariance matrix aligned along the scanning direction.
The results are shown in Figure \ref{fig:dir_noise}. Note that our network was {\em not} re-trained for this specific model.

\revised{In this setting, the non-data-driven methods, such as jet-fitting and edge-aware filtering, perform well since they are not specialized to any noise model. Even though our method was never trained on this type of noise, it still performs on par with the best methods on low to medium noise settings, and is only outperformed on high noise settings by non-data-driven methods with parameters tuned for the given noise strength.}

\subsubsection{Structured noise} \label{sssec:structnoise}
\begin{figure}[h!]
    \centering
    \includegraphics[width=\columnwidth]{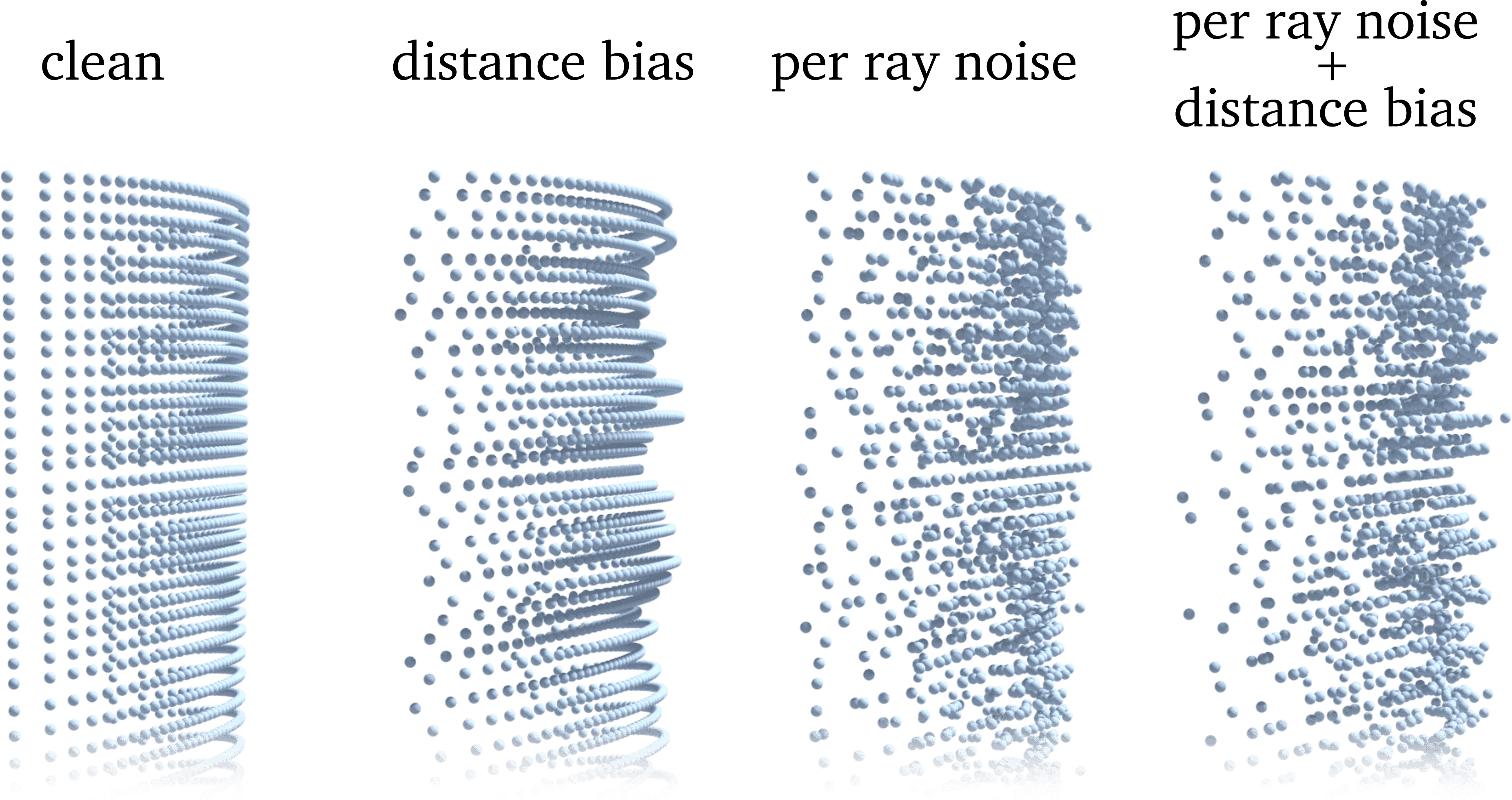}
    \caption{  Simulated noise of a Velodyne HDL-64E scanner. Here we show an example of the two noise types introduced by this scanner: distance bias (db) adds an error to the depth of scan lines, while ray noise (rn) adds an error to the depth of each point. Note that the points in the ground truth (left) are not distributed uniformly, so we use the distance of each point to the surface as error measure for experiments with this dataset.
    }
    \label{fig:lidar_noise_types}
\end{figure}

We   evaluated our method on a simulated LIDAR dataset (\velodyne) generated using BlenSor \cite{gschwandtner2011blensor}, which  models various types of range scanners. We chose to simulate a rotating LIDAR, in particular a Velodyne HDL-64E scanner. BlenSor implements two types of sensor specific error for this scanner: first, a distance bias for each laser unit; and second, a per-ray Gaussian noise. The different effects of the noise types can be observed in Figure~\ref{fig:lidar_noise_types}. We use the shapes in \name\ dataset for this experiment. After being normalized, each shape is scanned from $\theta \in [0^\circ,180^\circ]$ with distance bias with standard deviation 0\%, 0.5\% and 1\% and a per-ray noise of standard deviation 0\%, 0.5\% and 1\%.

Quantitative numbers are presented in Table~\ref{table:lidar}. Here, we use the distance to the surface instead of the Chamfer measure, since this type of scanner naturally produces non-uniform point clouds (even in the ground truth). In bold, we highlight the two best performing methods.
We evaluate two versions of our method: one version trained on the unstructured noise described in Section~\ref{sec:dataset}, and one version re-trained on the \velodyne\ dataset. The re-trained version significantly outperforms all other methods, while the non-retrained version still performs competitively.
In Figure~\ref{fig:qual_lidar}, we show a qualitative evaluation of our results compared to the two other best performing methods. We show the denoised scans for two shapes: cylinder and dragon, with distance bias (first two rows) and with distance bias and per-ray noise (last two rows). Both jet and bilateral methods preserve an amount of structured noise (see the cylinder shape). Jet medium produces artefact points in areas of high details especially on the dragon shape.

\begin{table}[t!]
\centering
\caption{\revised{Comparison with state-of-the-art methods on the \velodyne\ datasets. We show the root mean square distance-to-surface (RMSD) for each method. The first column evaluates the methods on a dataset with only distance bias as noise and the second column with added per-ray noise (see Figure~\ref{fig:lidar_noise_types} for examples of the noise types). PointProNets was re-trained on the dataset, and for our method we show both a re-trained version, and a version trained on the original dataset (Section~\ref{sec:dataset}).}
}
\begin{tabular}{l|r|r}
 & \velodyne\ (db) & \velodyne\ (db+rn) \\
\hline \hline
jet small & $5.46$& $5.78$ \\
jet medium & $4.91$&  $5.18$\\
jet large &$9.68$ & $9.67$\\
edge-aware small & $5.50$& $6.36$\\
edge-aware med. &$5.48$ & $5.77$\\
edge-aware large & $11.31$& $11.53$\\
bilateral & $\textbf{4.53}$& $\textbf{4.99}$ \\
PointProNets & $17.47$ & $22.02$ \\
ours & $5.83$ & $7.03$ \\

\textbf{ours re-trained} & $\textbf{4.07}$& $\textbf{4.27}$\\
\end{tabular}
\label{table:lidar}
\end{table}

\begin{figure*}[t!]
    \centering
    \includegraphics[width=\textwidth]{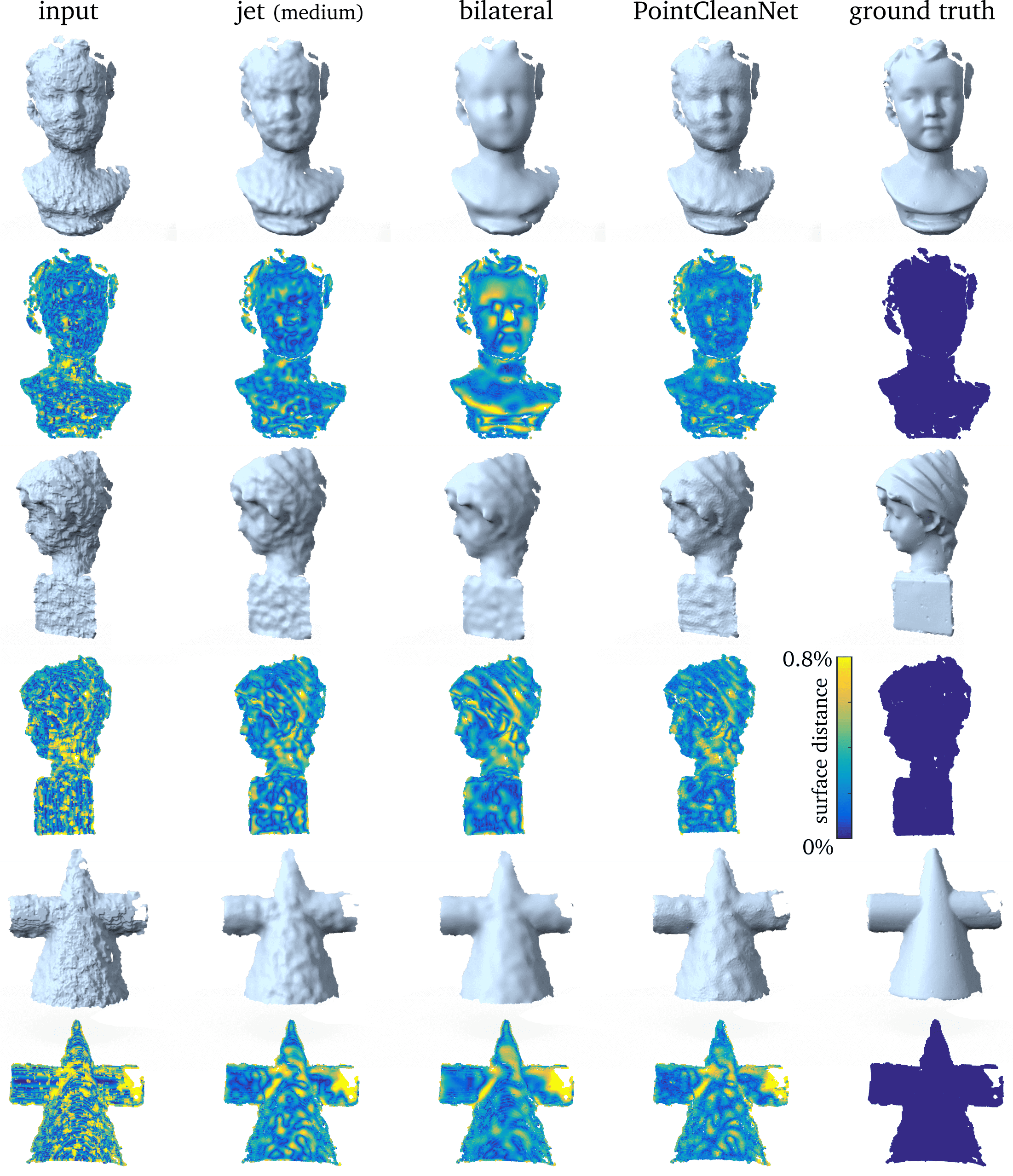}
    \caption{\revised{Qualitative comparison of the three best performing methods on the \kinectone\ dataset. For each shape, we compare a Poisson reconstruction and the normalized distance to the ground truth surface of the denoised point sets computed by each method.}}
    \label{fig:qual_kinectv1}
\end{figure*}

Table~\ref{table:kinectv1} and Figure~\ref{fig:qual_kinectv1} evaluate our model on the \kinectone\ dataset introduced in \cite{wang2016mesh}.  We trained the model on the scans of the shapes David, big-girl and pyramid from this dataset, and tested on a subset of the scans from boy, girl and cone with a radius of 2.5\% of the shapes' bounding box diagonals. Our trained network achieves the best performance. In Figure~\ref{fig:qual_kinectv1},  we show the Poisson reconstructions \cite{Kazhdan:2013:SPS:2487228.2487237} and distances from the ground truth for the three best performing methods: jet medium, bilateral and ours. The normals where computed using the normal estimation tool from Meshlab. Note that jet-fitting method tends to preserve the structured noise while the bilateral method tends to oversmooth shape.
We also re-trained \name\ on the \kinecttwo\ dataset described in \cite{wang2016mesh}. As shown in Table~\ref{table:kinectv1} our method performs well compared to other methods.

\begin{table}[b!]
\centering
\caption{
\revised{Comparison with state-of-the-art methods on the \kinectone\  and \kinecttwo\ datasets. We show the chamfer measure for each method. PointProNets was re-trained on each dataset, and for our method we show both a re-trained version, and a version trained on the original dataset (Section~\ref{sec:dataset}).}}

\begin{tabular}{l|r|r}
 & \kinectone\ & \kinecttwo\ \\
\hline \hline
jet small & $5.10$ & $6.36$\\
jet medium & $\textbf{4.69}$ & $\textbf{6.16}$\\
jet large & $5.40$ & $8.63$\\
edge-aware small & $5.21$ & $6.38$\\
edge-aware med. & $4.78$ & $6.53$\\
edge-aware large & $6.85$ & $13.10$\\
bilateral & $4.72$ & $\textbf{6.04}$ \\
PointProNets & $7.39$ & $12.81$\\
ours & $5.02$  & $6.42$\\
\textbf{ours re-trained} & $\textbf{4.57}$ &  $6.26$\\
\end{tabular}
\label{table:kinectv1}
\end{table}

\begin{figure}[h!]
    \centering
    \includegraphics[width=\columnwidth]{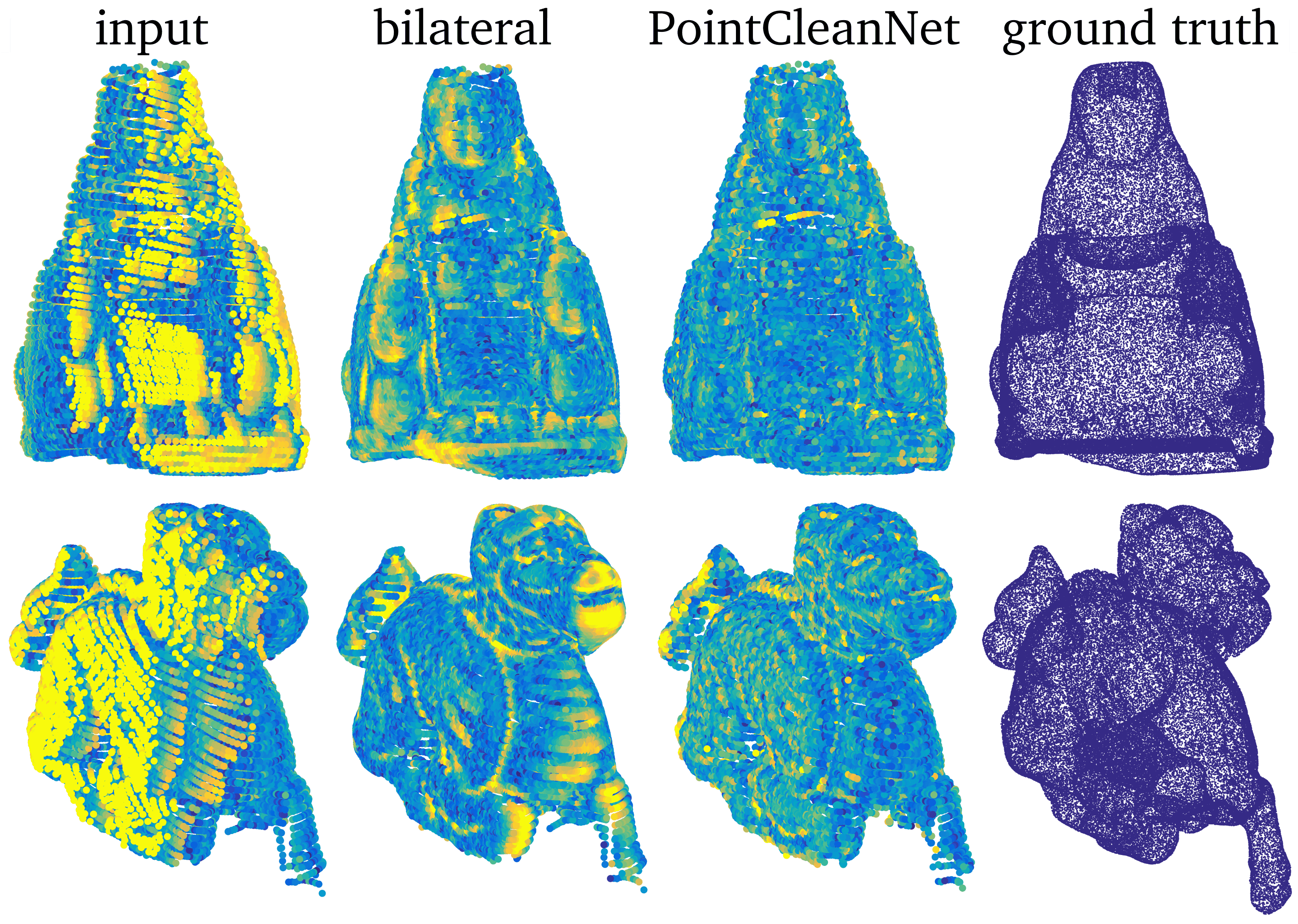}
    \caption{ \minorrevision{Qualitative comparison with bilateral filtering \cite{digne2017bilateral} on the misaligned scans dataset. We show the normalized distance to the ground truth surface.\vspace{-2mm}}} \label{fig:misaligned}
\end{figure}

\minorrevision{
\subsubsection{Misaligned data}
\label{sec:misaligned_data}
 
We evaluated our method on a dataset containing misaligned scans. For this, we generate scans of each mesh from the \name\ dataset by simulating a rotating Velodyne HDL-64E scanner using BlenSor and scan each shape from 6 equally spaced angles between $0^\circ$ and $360^\circ$ without noise. To simulate the artefacts of misaligned scans, we apply a rotation centered at the scanner position and of angle uniformly sampled in $[-\theta, \theta]$ to each scan. Our misaligned scans dataset was built with $\theta$ being valued at  $1^\circ$ and $2^\circ$. Figure~\ref{fig:cross_section} shows a cross section of two shapes from the misaligned scans dataset.

\begin{figure}[h!]
    \centering
    \includegraphics[width=\columnwidth]{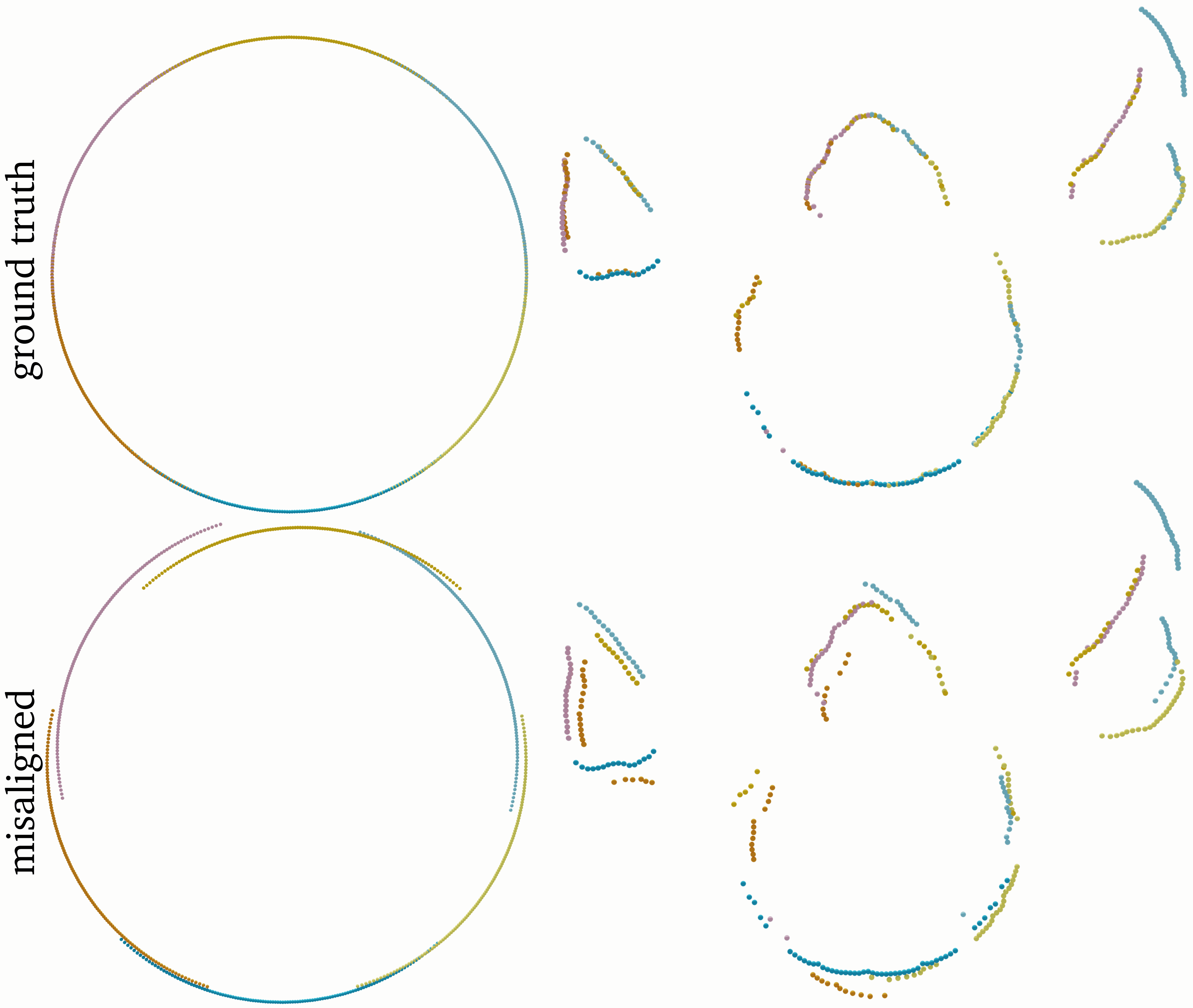}
    \caption{ \minorrevision{Cross section of cylinder  (left) and armadillo (right) misaligned scans with $\theta =2^\circ$. Each color corresponds to a scan from a single angle. 
    }}
    \label{fig:cross_section}
  \end{figure}

We present a quantitative evaluation in Table~\ref{table:misaligned}. Similar to the \velodyne\ dataset, we only evaluate the distance to the surface, since the produced scans are not uniform. We evaluate a model that we re-trained on this dataset, as well as the model that was only trained on the basic \name\ dataset. We compare both to state-of-the-art methods. Note that our re-trained model has the best performance. In Figure~\ref{fig:misaligned} we also show a qualitative comparison of our method with the bilateral filtering method \cite{digne2017bilateral}. We observe that \name\ is able to merge different scans while bilateral filtering keeps some artefacts from the misaligned scans (first row) and even introduces noise (second row). 

}

\begin{table}[b!]
\centering
\caption{\minorrevision{ Comparison with state-of-the-art methods on the misaligned scans dataset. We report the root mean square distance-to-surface (RMSD). We consider both a re-trained version and a version trained on the original dataset of \name\ . }}
\begin{tabular}{l|r}

 & misaligned scans \\
\hline \hline
jet small & $5.10$\\
jet medium &  $\textbf{4.74}$\\
jet large & $7.52$\\
edge-aware small & $5.43$ \\
edge-aware med. & $4.85$\\
edge-aware large & $6.62$\\
bilateral & $5.08$ \\
ours & $5.23$\\
\textbf{ours re-trained} & $\textbf{4.35}$ \\
\end{tabular}
\label{table:misaligned}
\end{table}

\subsubsection{Generalization to real-world data.}

Figure~\ref{fig:teaser} and the first two results in Figure~\ref{fig:results_plate}~(left) show the result of our approach on real data obtained with the plane swift algorithm~\cite{wolff2016point}, an image-based 3D reconstruction technique. Statue, torch, and scarecrow input point clouds each contain 1.4M points. Since in this case no ground truth is available, we only show the qualitative results obtained using our method. Note that although trained on an entirely different dataset, \name\ still produces high quality results on this challenging real-world data.
The next three shapes in Figure \ref{fig:results_plate} show results on other external raw point clouds, while the last one shows a shape with sharp edges.

\section{Conclusion, Limitations and Future Work}

We presented \name, a learning-based framework  that consumes noisy point clouds and outputs clean ones by removing outliers and denoising the remaining points with a displacement back to the underlying (unknown) scanned surface. 
One key advantage of the proposed setup is the simplicity of using the framework at test time as it neither requires additional parameters nor noise/device specifications from the user. In our extensive evaluation, we demonstrated that \name\ consistently  outperforms state-of-the-art denoising approaches (that were provided with manually tuned parameters) on a range of models under various noise settings. 
Given its universality and ease of use, \name\ can be readily integrated with any geometry processing workflow that consumes raw point clouds. \revised{Note that in our current framework, we still need paired noisy-clean data to train \name. An exciting future direction would be learn denoising directly from unpaired data. As a supervised learning method, our approach is also unlikely to succeed when noise characteristics during training are very different from the ones of the test data.}

While we presented a first learning architecture to clean raw point clouds directly, several future directions remain to be explored: (i)~First, as a simple extension, we would like to combine the outlier removal and denoising into a single network, rather than two separate parts. (ii)~Further, to increase efficiency, we would like to investigate how to perform denoising at a patch-level rather than per-point level. This would require designing a scheme to combine denoising results from overlapping patches. 
(iii)~Although \name\ already produces a uniform point distribution on the underlying surface if the noisy points are uniformly distributed, we would like to investigate the effect of a specific uniformity term in the loss function (similar to \cite{Yin:2018:PBP:3197517.3201288}) to also produce a uniform distribution for non-uniform noisy points. The challenge, however, would be to restrain the points to remain on the surface and not deviate off the underlying surface. %
(iv)~Additionally, it would be interesting to investigate how to allow the network to upsample points, especially in regions with insufficient number of points, or to combine it with existing upsampling methods such as \cite{yu2018pu} This would be akin to the `point spray' function in more traditional point cloud processing toolboxes. 
(v)~Finally, we would like to investigate how to train a point cloud cleanup network {\em without} requiring paired noisy-clean point clouds in the training set. If successful, this will enable directly handling noisy point clouds from arbitrary scanning setups without requiring explicit noise model or examples of denoised point clouds at training time. We plan to draw inspiration from related \textit{unpaired} image-translation tasks where generative adversarial setups that have been successfully used.

\section{Acknowledgements}

Parts of this work were supported by the KAUST OSR Award No. CRG-2017-3426, a gift from the NVIDIA Corporation, the ERC Starting Grants EXPROTEA (StG-2017-758800) and SmartGeometry (StG-2013-335373), a Google Faculty Award, and gifts from Adobe.


\bibliographystyle{eg-alpha-doi}

\bibliography{paper}

\begin{figure*}[t!]
    \centering
    \includegraphics[width=0.9\textwidth]{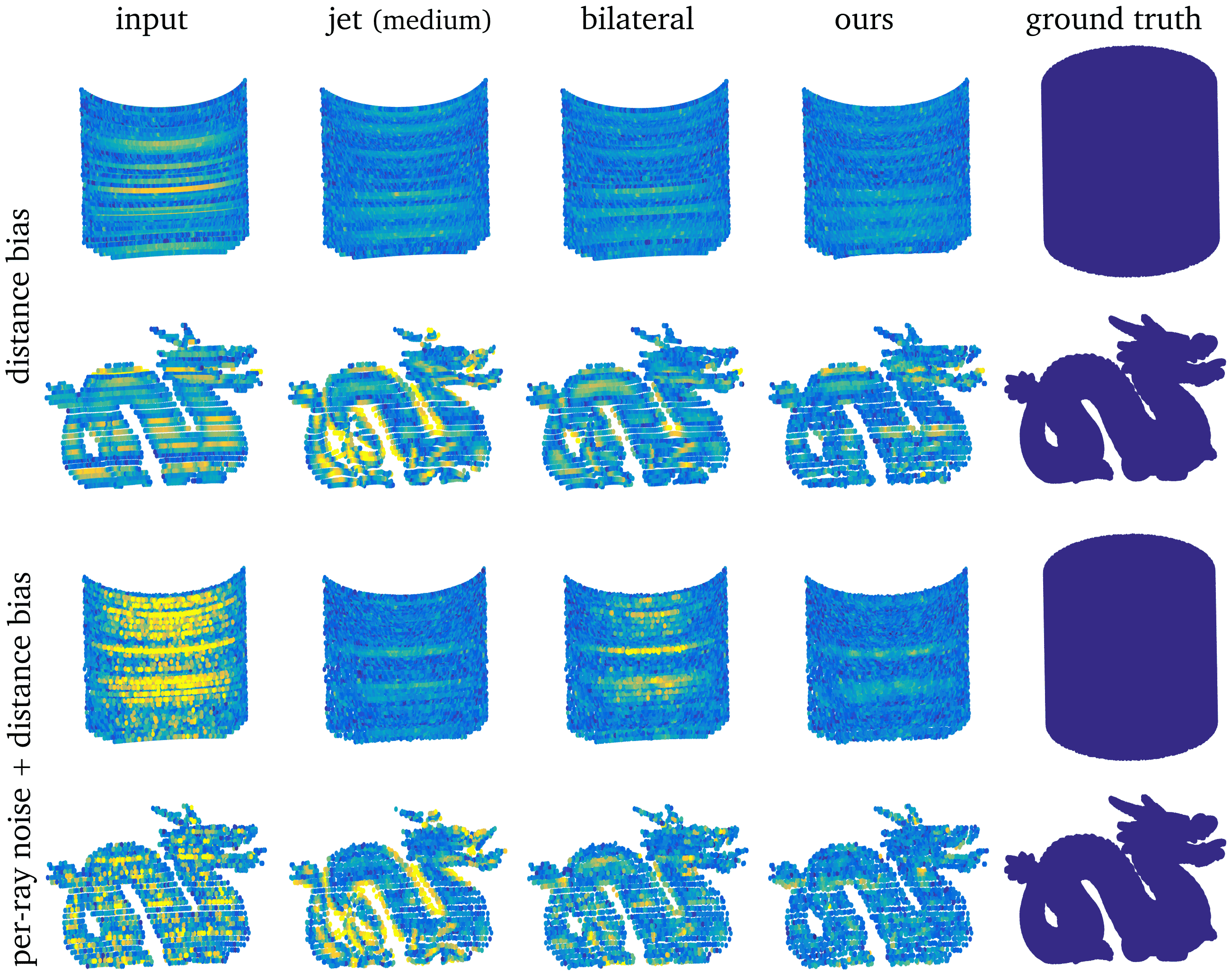}
    \caption{\revised{Qualitative comparison with state-of-the-art methods on the \velodyne\ dataset. We display the normalized distance to the ground truth surface. The two top rows are evaluated on a dataset with only distance bias as noise and the two bottom rows with added per-ray noise. The simulated scanner noise has a high spatial correlation along the horizontal scan-lines, and lower correlation vertically across scan-lines. In this setting, jet fitting introduces significant error in detailed surface regions, while bilateral denoising has high residual error in the examples that have both noise types. \name\ successfully learns the noise model, resulting in lower residual error.}}
    \label{fig:qual_lidar}
\end{figure*}

\appendix 

\section{}

Figure~\ref{fig:qual_lidar} shows a qualitative comparison with state-of-the-art methods on the \velodyne\ dataset described in Section \ref{sssec:structnoise}.


\end{document}